\documentclass[sigconf]{aamas} 
\usepackage{algorithmic}
\usepackage{algorithm}
\usepackage{listings}

\usepackage{balance} 

\setcopyright{ifaamas}
\acmConference[AAMAS '24]{Proc.\@ of the 23rd International Conference
on Autonomous Agents and Multiagent Systems (AAMAS 2024)}{May 6 -- 10, 2024}
{Auckland, New Zealand}{N.~Alechina, V.~Dignum, M.~Dastani, J.S.~Sichman (eds.)}
\copyrightyear{2024}
\acmYear{2024}
\acmDOI{}
\acmPrice{}
\acmISBN{}

\acmSubmissionID{79}

\title[AAMAS-2024 Formatting Instructions]{TaxAI: A Dynamic Economic Simulator and Benchmark for Multi-Agent Reinforcement Learning}

\author{Qirui Mi}
\affiliation{
  \institution{Institute of Automation, CAS}
  \institution{School of Artificial Intelligence, UCAS}
  \city{Beijing}
  \country{China}}
\email{miqirui2021@ia.ac.cn}

\author{Siyu Xia}
\affiliation{
  \institution{Institute of Automation, CAS}
  \institution{School of Artificial Intelligence, UCAS}
  \city{Beijing}
  \country{China}
  }
\email{xiasiyu2023@ia.ac.cn}

\author{Yan Song}
\affiliation{
  \institution{Institute of Automation, CAS}
  \city{Beijing}
  \country{China}}
\email{yan.song@ia.ac.cn}

\author{Haifeng Zhang*}\thanks{*Corresponding to Haifeng Zhang $\langle$\href{haifeng.zhang@ia.ac.cn}{haifeng.zhang@ia.ac.cn}$\rangle$. }
\affiliation{
  \institution{Institute of Automation, CAS}
  \institution{School of Artificial Intelligence, UCAS}
  \institution{Nanjing Artificial Intelligence Research of IA}
  \city{Beijing}
  \country{China}}
\email{haifeng.zhang@ia.ac.cn}

\author{Shenghao Zhu}
\affiliation{
  \institution{University of International \\Business and Economics}
  \city{Beijing}
  \country{China}}
\email{zhushenghao@yahoo.com}

\author{Jun Wang}
\affiliation{
  \institution{University College London}
  \city{London}
  \country{United Kingdom}}
\email{jun.wang@cs.ucl.ac.uk}

\begin{abstract}
Taxation and government spending are crucial tools for governments to promote economic growth and maintain social equity. However, the difficulty in accurately predicting the dynamic strategies of diverse self-interested households presents a challenge for governments to implement effective tax policies. Given its proficiency in modeling other agents in partially observable environments and adaptively learning to find optimal policies, Multi-Agent Reinforcement Learning (MARL) is highly suitable for solving dynamic games between the government and numerous households. Although MARL shows more potential than traditional methods such as the genetic algorithm and dynamic programming, there is a lack of large-scale multi-agent reinforcement learning economic simulators. Therefore, we propose a MARL environment, named \textbf{TaxAI}, for dynamic games involving $N$ households, government, firms, and financial intermediaries based on the Bewley-Aiyagari economic model. Our study benchmarks 2 traditional economic methods with 7 MARL methods on TaxAI, demonstrating the effectiveness and superiority of MARL algorithms. Moreover, TaxAI's scalability in simulating dynamic interactions between the government and 10,000 households, coupled with real-data calibration, grants it a substantial improvement in scale and reality over existing simulators. 
Therefore, TaxAI is the most realistic economic simulator for optimal tax policy, which aims to generate feasible recommendations for governments and individuals.
\end{abstract}


\newcommand{\BibTeX}{\rm B\kern-.05em{\sc i\kern-.025em b}\kern-.08em\TeX}

\makeatletter
\gdef\@copyrightpermission{
	\begin{minipage}{0.3\columnwidth}
		\href{https://creativecommons.org/licenses/by/4.0/}{\includegraphics[width=0.90\textwidth]{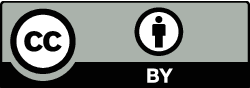}}
	\end{minipage}\hfill
	\begin{minipage}{0.7\columnwidth}
		\href{https://creativecommons.org/licenses/by/4.0/}{This work is licensed under a Creative Commons Attribution International 4.0 License.}
	\end{minipage}
	\vspace{5pt}
}
\makeatother

\begin{document}


\pagestyle{fancy}
\fancyhead{}


\maketitle 

\section{Introduction}
The invisible hand~\cite{smith2010theory,smith2023inquiry} of the market is not omnipotent, and in reality, all countries rely on government intervention to promote economic development and maintain social fairness. The extent of government intervention varies from country to country, such as a free market economy~\cite{sunstein1997free, baumol2002free}, planned economies~\cite{mcmillan1992reform, peng1996growth} or mixed economies~\cite{nee1992organizational, ikeda2002dynamics, johnson2014mixed}. However, determining the optimal government intervention degree is challenging for several reasons. Firstly, extracting relevant and actionable information from the complex society is arduous. Secondly, governments face difficulties in effectively modeling a vast and heterogeneous population with diverse preferences and characteristics. Lastly, the behavioral response of individuals to incentives remains highly unpredictable.

In this intricate matter of government intervention, we opt to investigate a crucial and efficacious tool, \textbf{tax policy}, which is commonly studied using agent-based modeling (ABM)~\cite{davidssonAgentBasedSocial2002,dingSimtorealTransferOptical2020} in economics. ABM is an effective approach to simulate individual behaviors and show the relationship between micro-level decisions and macro-level phenomena. However, traditional ABM suffers from simplicity and subjectivity in setting model parameters and behavior rules, making it difficult to simulate realistic scenarios~\cite{liAgentBasedSocialSimulation2008,dingSimtorealTransferOptical2020,shiArtificialIntelligenceSocial2020}. 
While Multi-Agent Reinforcement Learning (MARL) surpasses traditional ABM settings by offering optimal actions based on evolving state information. Some MARL algorithms perform well in partially observable environments and adaptively learn to reach equilibrium solutions~\cite{foerster2016learning, oliehoek2016concise, sukhbaatar2016learning}. Hence, MARL is well-suited for addressing dynamic game problems involving the government and a large population.
However, despite the significant advantages of MARL, there is currently a shortage of large-scale MARL economic simulators designed specifically for the study of tax policies.
In the existing literature, the AI Economist~\cite{zhengAIEconomistTaxation2022} and the RBC model~\cite{curryAnalyzingMicroFoundedGeneral2022} emerge as the most closely related simulators to TaxAI. However, these models exhibit certain limitations, notably a partial grounding in economic theory, limited scalability in simulating a significant number of agents, and the absence of calibration using real-world data (as detailed in Table~\ref{table: comparison}). 
Therefore, proposing a more realistic MARL environment to study optimal tax policies and solve dynamic games between the government and the population holds significant research and practical value.

\begin{figure*}[h]
    \centering
    \includegraphics[scale=0.55]{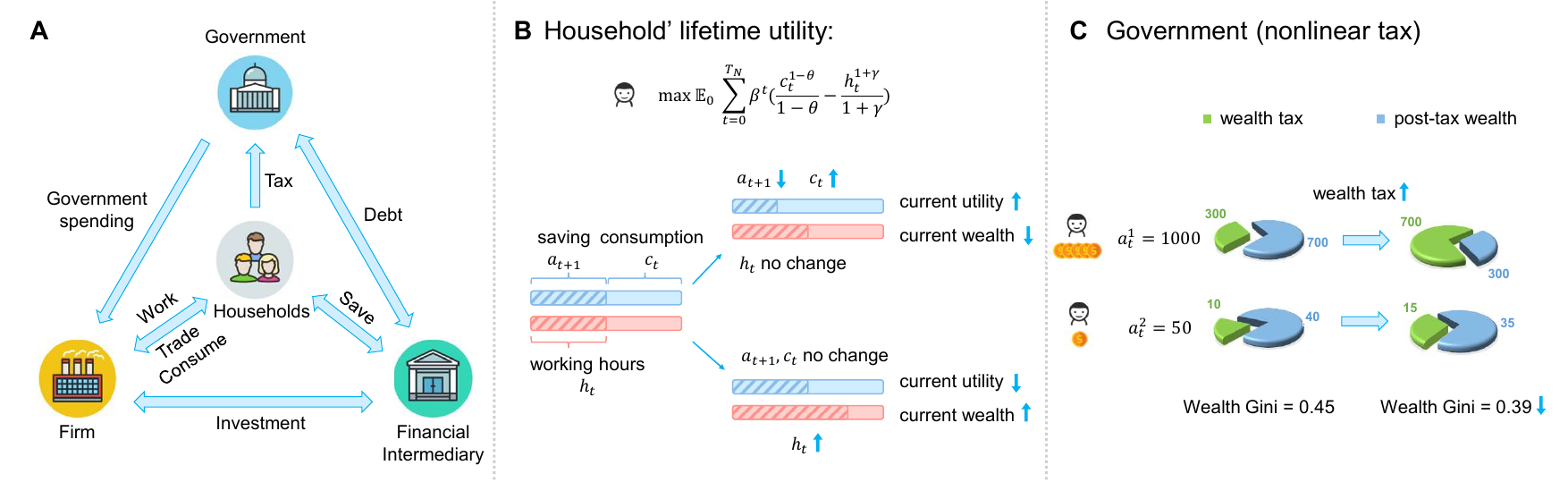}
    \caption{Model Dynamics in the Bewley-Aiyagari Model. \textbf{A:} Economic activities among the government, the firm, the financial intermediary, and households.  \textbf{B:} The influence of households' saving and labor strategies on current utility and wealth. Households must strike a balance between consumption and savings, as well as work and leisure, to optimize lifetime utility. Increasing consumption enhances current utility but reduces current wealth, affecting future utility. Longer working hours yield higher labor income, thereby increasing wealth, but simultaneously result in disutility. \textbf{C:} The effect of government taxation on households' wealth. The social planner employs a nonlinear taxation, applying varying tax rates based on different assets. As tax rates rise, the taxes paid by households increase, with wealthier household contributing more. This narrows the gap in households' post-tax wealth, leading to a reduction in the Gini coefficient for wealth distribution.}
    \label{fig:dynamics}
\end{figure*}

\begin{table}[t]
	\caption{A comparison of MARL simulators for optimal taxation problems.}
	\label{table: comparison}
	\begin{tabular}{lccc}\toprule
    {Simulator} & {AI} & {RBC} & {\textbf{TaxAI}} \\
    {} & {Economist} & {Model} & {(ours)} \\
    \midrule
    {Households'} & {10} &{100} &{10000}\\
    {Maximum Number} & {} &{} &{}\\
    {Tax Schedule} & {Non-linear} & {Linear} & {Non-linear} \\
    {Tax Type} & {Income} & {Income} & {Income\&}\\
    {} & {} & {} & {Wealth\&} \\
    {} & {} & {} & {Consumption } \\
    {Social Roles' Types} & {2} & {3} &{4}\\
    {Saving Strategy} & {$\times$} & {\checkmark}& {\checkmark} \\
    {Heterogenous Agent} & {\checkmark} & {\checkmark}& {\checkmark} \\ 
    {Real-data Calibration} & {$\times$} & {$\times$} & {\checkmark} \\
    {Open source} & {\checkmark} &{$\times$} & {\checkmark} \\
    {MARL Benchmark} & {$\times$} & {$\times$} & {\checkmark} \\
    \bottomrule 
	\end{tabular}
\end{table}
Therefore, we introduce a dynamic economic simulator, TaxAI, based on the Bewley-Aiyagari economic model~\cite{aiyagariOptimalCapitalIncome1995, aiyagariUninsuredIdiosyncraticRisk1994}, which is widely used to study capital market frictions, wealth distribution, economic growth issues. By incorporating the Bewley-Aiyagari model, TaxAI benefits from robust theoretical foundations in economics and models a broader range of social roles (shown in Figure~\ref{fig:dynamics}).
Based on TaxAI, we benchmark 2 economic methods and 7 MARL algorithms, optimizing fiscal policy for the government, working and saving strategies for heterogeneous households. In our experiments, we compared 9 baselines across four distinct tasks, evaluating them from both macroeconomic and microeconomic perspectives. Our results reveal the tax-avoidance behavior of MARL-based households and the varying saving and working strategies among households with different levels of wealth. Finally, we test the TaxAI environment using 9 baselines with households' number ranging from 10, 100, 1000, and even up to 10,000, demonstrating its capability to simulate large-scale agents.
In summary, our work encompasses the following three contributions:

\textbf{1. A dynamic economic simulator TaxAI}. The simulator incorporates multiple roles and main economic activities, employs real-data calibration, and facilitates simulations of up to 10,000 agents.
These features provide a more comprehensive and realistic simulation than existing simulators.

\textbf{2. Validation of MARL feasibility in optimizing tax policies}. We implemented 2 traditional economic approaches and 7 MARL methods to solve optimal taxation for the social planner, and optimal saving and working strategies for households. The results obtained through MARL methods surpass those achieved by traditional methods.

\textbf{3. Economic analysis of different policies}. We conducted assessments from both macroeconomic and microeconomic perspectives, uncovering tax-avoidance behaviors among MARL-based households in their pursuit of maximum utility. Furthermore, we observed distinct strategies among households with differing levels of wealth.

Codes for the TaxAI simulator and algorithms are shown in the GitHub repository~\href{https://github.com/jidiai/TaxAI}{https://github.com/jidiai/TaxAI}.

\section{Related Works}\label{related_work}

\paragraph{Classic Tax Models}
Economic models provide powerful tools for modeling economic activities and explaining economic phenomena. In microeconomics, the Supply and Demand model~\cite{smith1937wealth, gale1955law} reveals the mechanism behind market price formation, while the marginal utility theory~\cite{kauder2015history} underscores the significance of consumption decisions. In macroeconomics, the Keynesian Aggregate Demand-Aggregate Supply Model~\cite{barro1994aggregate, dutt1996keynesian} addresses short-term fluctuations and policy effects~\cite{dutt2006aggregate}. The Comparative Advantage Theory~\cite{hunt1995comparative, costinot2009elementary} in international trade explains collaborations across nations. The Quantity Theory of Money~\cite{lucas1980two, friedman1989quantity} investigates the relationship between money supply and price levels.
Regarding the optimal tax problem, the Ramsey-Cass-Koopmans (RCK) model~\cite{cass1965optimum, koopmans1963concept} studies the consumption and savings decisions of representative agents but ignoring individual heterogeneity. The Diamond-Mirrlees model~\cite{diamond1971optimala, diamond1971optimalb} considers the role of taxes and labor supply in social welfare but overlooks income and asset taxes. The Overlapping Generations (OLG) model~\cite{samuelson1958exact} emphasizes intergenerational inheritance and resource transfers~\cite{diamond1965national, galor1992two}. In contrast, the Bewley-Aiyagari model~\cite{aiyagari1995optimal, bewley1986stationary} can assess the impact of taxation on growth, wealth distribution and welfare while simulating real-world income disparities and risk-bearing capacity of individuals. This makes the Bewley-Aiyagari model an ideal choice for studying optimal taxation and household strategies. 

\vspace{-0.4em}
\paragraph{Traditional Economic Methods} 
The optimal tax policy and wealth distribution~\cite{benhabib2018skewed} have been extensively studied in economics. Existed works~\cite{aiyagari1995optimal,chari1999optimal} have utilized mathematical programming methods to address decision-making processes related to governments and households~\cite{boar2022efficient,bakics2015transitional}. However, these approaches oversimplify decision-makers rationality and fail to consider autonomous learning abilities and environmental uncertainties. In contrast, dynamic programming-based approaches~\cite{domeij2004distributional,dyrda2016optimal} consider long-term consequences and environmental dynamics but struggle to model non-rational behaviors~\cite{carroll2023optimal}. Alternative approaches, such as empirical rules~\cite{love2013optimal,heathcote2017optimal} and Agent-Based Modeling (ABM)~\cite{tesfatsion2001introduction, steinbacher2021advances}, have emerged to address these limitations. ABM enables the exploration of micro-level behaviors and their impact on macro-level phenomena, showing in the ASPEN model~\cite{basu1998aspen}, income distribution~\cite{dosi2013income} and transaction development~\cite{klos2001agent}. Despite the abundance of research in economics based on ABM, this approach often involves relatively simplistic and subjective specifications of individual behavior, making it challenging to investigate the dynamic optimization of individual strategies.

\vspace{-0.4em}
\paragraph{MARL and Simulators for Economy} 
MARL aims to address issues of cooperation~\cite{busoniuMultiagentReinforcementLearning2010} and competition~\cite{zhangMultiagentReinforcementLearning2021} among multiple decision-makers~\cite{suttonReinforcementLearningIntroduction2018}. The simplest MARL method Independent Learning~\cite{tanMultiagentReinforcementLearning1993,maes1990learning}, including IPPO~\cite{de2020independent}, and IDDPG~\cite{lillicrap2015continuous}, involves each agent learning and making decisions independently, disregarding the presence of other agents~\cite{tampuuMultiagentCooperationCompetition2015}. In the Centralized Training Decentralized Execution (CTDE) algorithms, like MADDPG~\cite{loweMultiAgentActorCriticMixed2017}, QMIX~\cite{rashidMonotonicValueFunction2020}, and MAPPO~\cite{yuSurprisingEffectivenessPpo2022}, agents share information during training but make decentralized decisions during execution to enhance collaborative performance. 
To address significant computational and communication overhead posed by a growing number of agents~\cite{yangMeanFieldMultiagent2018a}, Mean-Field Multi-Agent Reinforcement Learning (MF-MARL)~\cite{zhouFactorizedQlearningLargescale2019} simplifies the problem by treating homogeneous agents as distributed particles~\cite{guMeanfieldMultiagentReinforcement2021}.
On the other hand, Heterogeneous Agent Reinforcement Learning (HARL)~\cite{zhong2023heterogeneous}, HAPPO and HATRPO, is designed to achieve effective cooperation in a general setting involving heterogeneous agents.

Currently, there are not many efforts employing MARL methods to determine optimal tax policies and individual strategies. The closest works to our paper include AI economist~\cite{zhengAIEconomistTaxation2022} and the RBC model~\cite{curryAnalyzingMicroFoundedGeneral2022}. While they both account for fundamental economic activities, they lack large-scale agent simulation, real-data calibration, and MARL benchmarks. These limitations make practical implementation challenging, which is why we introduce TaxAI.
Besides, prior research has already explored reinforcement learning-based approaches in some subproblems. 
For instance, in addressing optimal savings and consumption problems, some studies~\cite{shiCanAIAgent2021, ruiLearningZeroHow2022, atashbarAIMacroeconomicModeling2023} have utilized single-agent RL to model the representative agent or a continuum of agents. Meanwhile, others have employed MARL to solve rational expectations equilibrium~\cite{kurikshaEconomyNeuralNetworks2021, hillSolvingHeterogeneousGeneral2021}, optimal asset allocation and savings strategies~\cite{ozhamaratliDeepReinforcementLearning2022a}. 
Regarding optimal government intervention problems, research has explored the application of RL in investigating optimal monetary policy~\cite{hinterlangOptimalMonetaryPolicy2021, chenDeepReinforcementLearning2023}, market prices~\cite{danassis2023ai, schlechtinger2023price}, international trade~\cite{schIntelligenceEconomyEmergent2021}, redistribution systems~\cite{kosterHumancentredMechanismDesign2022, yamanEmergenceDivisionLabor2022}, and the cooperative relationship between central and local governments under COVID-19~\cite{trottBuildingFoundationDataDriven2021}.

\section{Bewley-Aiyagari Model}\label{BA model}
In comparison to classical economic models in Section~\ref{related_work}, we contend that the Bewley-Aiyagari model serves as the most suitable theoretical foundation for investigating optimal tax policy for the government and optimal savings and labor strategies for households. In this section, we provide a brief overview of the Bewley-Aiyagari model. This model encompasses four key societal roles: N households, a representative firm, a financial intermediary, and the government. The interactions among them are illustrated in Figure~\ref{fig:dynamics}. All model variables and their corresponding symbols are organized in Appendix Table~\ref{table:model parameter}. More details on model assumptions and dynamics are shown in Appendix~\ref{assumption}, \ref{dynamics}.

\subsection{$N$ Households}
To avoid differences in age, gender, and personality, individuals are modeled as households, whose main activities include production, consumption, saving, and tax payments.
At timestep $t$, households' income $i_t$ is derived from two sources. They \textbf{work} in the technology firm for the labor income $W_t e_t h_t$, which depends on the wage rate $W_t$, the individual labor productivity levels $e_t$ and the working hours $h_t$. On the other hand, households can only engage in \textbf{savings} and are assumed not to borrow ($a_{t} \geq 0$). They earn interest income $r_{t-1} a_t$ from savings, which depends on household asset $a_t$ and the return to savings $r_{t-1}$. 
\begin{equation}
    i_t = W_t e_t h_t + r_{t-1} a_t
\end{equation}
In this model, $N$ households are heterogeneous in terms of labor productivity levels $e_t$ and initial asset $a_0$, and we model $e_t$ as either being in a super-star ability or a normal state. In the normal state, it follows an AR(1) process~\cite{boar2022efficient}, a model commonly used for analyzing and forecasting time series data.
\begin{equation}
    \log e_t=\rho_e \log e_{t-1}+\sigma_e u_t
\end{equation}
where $\rho_e$ is the persistence and $\sigma_e$ is the volatility of the standard normal shocks $u_t$. 
In the super-star state, the labor market ability is $\bar{e}$ times higher than the average. The transition of households from the normal state to the super-star state occurs with a constant probability $p$ while remaining in the super-star state has a constant probability $q$.

In addition, each household seeks to maximize lifetime utility (\ref{households_utility}) depends on \textbf{consumption} $c_t$ and \textbf{working hours} $h_t$, subject to budget constraint, where $\beta$ is the discount factor, $\theta$ is the coefficient of relative risk aversion (CRRA), and $\gamma$ represents the inverse Frisch elasticity. $T_N$ denotes the maximum steps.
\begin{equation}\label{households_utility}
    \begin{aligned}
        & \max \quad \mathbb{E}_0 \sum_{t=0}^{T_N} \beta^t\left(\frac{c_t^{1-\theta}}{1-\theta}-\frac{h_t^{1+\gamma}}{1+\gamma}\right) \\
        \text{s.t. } & (1+\tau_s)c_t+a_{t+1}=i_t-T(i_t)+a_t-T^a(a_t)
    \end{aligned}
\end{equation}

The households are required to \textbf{pay taxes} to the government, including consumption taxes $\tau_s$, income taxes $T(i_t)$, and asset taxes $T^a(a_t)$, the last two are expressed by a nonlinear $\mathrm{HSV}$ tax function~\cite{heathcote2017optimal,benabou2002tax},
\begin{equation}
    T\left(i_t\right)=i_t-(1-\tau) \frac{i_t^{1-\xi}}{1-\xi},  \quad T^a\left(a_t\right)=a_t-\frac{1-\tau_a}{1-\xi_a} a_t^{1-\xi_a}
\end{equation} 
where $\tau, \tau_a$ determine the average level of the marginal income and asset tax, and $\xi, \xi_a$ determine the slope of the marginal income and asset tax schedule. It presents a free market economy when all taxes are equal to 0.

\subsection{Technology Firm}
\label{firms}
As the representative of all firms and industries, the firm converts capital and labor into goods and services. We assume it produces a homogeneous good with technology, which can meet the consumption need of households, following the Cobb–Douglas production function,
\begin{equation}
    Y_t=K_t^\alpha L_t^{1-\alpha}
\end{equation}
where $K_t$ and $L_t$ are capital and labor used for production, $\alpha$ is capital elasticity, and we normalize the output price to 1. The firm rent capital at a rental rate $R_t$ and hires labor at a wage rate $W_t$.
The produced output is used for all households' gross consumption $C_t$, government spending $G_t$, and physical capital investment $X_t=K_{t+1}-(1-\delta) K_t$, with the depreciation rate $\delta$, so the aggregate resource constraint is
\begin{equation}\label{gdp}
    Y_t = C_t+X_t+G_t
\end{equation}
Suppose the firm takes the marginal income from labor as households' wage rate $W_t$ and the marginal income from capital as the rental rate $R_t$.
\begin{equation}
    \begin{split}
            W_t = \frac{\partial Y_t}{\partial L_t} 
            = (1-\alpha) (\frac{K_t}{L_t})^\alpha, \quad R_t = \frac{\partial Y_t}{\partial K_t} = \alpha (\frac{K_t}{L_t})^{\alpha-1}
    \end{split}
\end{equation}
Market clearing on labor and goods is an important assumption for simplification, which means there is an equilibrium between supply and demand. The goods market clears by Walras' Law, and the labor market clearing condition is $L_t=\sum_i^N e_t^i h_t^i$.
 
\subsection{Government}
The government has multiple goals, such as promoting economic growth, maintaining social fairness and stability, and maximizing social welfare. To optimize these objectives, the government typically employs three tools, including government spending $G_t$, taxation $T_t$, and debt $B_t$ with the interest rate $r_{t-1}$.
For instance, when maximizing economic growth, the government's objective and the budget constraint are as follows:
\begin{equation}
    \begin{aligned}
        &\max \quad J = \mathbb{E}_0 \sum_{t=0}^{T_N} \beta^t \left(\frac{Y_t - Y_{t-1}}{Y_{t-1}}\right)\\
        & \text{s.t.} \left(1+r_{t-1}\right) B_t+G_t=B_{t+1}+T_t
    \end{aligned}
\end{equation}
where taxes $T_t$ are derived from personal income taxes, wealth taxes, and consumption taxes.
\begin{equation}
    T_t = \sum_i^N \left( T(i^i_t) + T(a^i_t) + \tau_s c_t^i \right)
\end{equation}
In addition to the task of maximizing social welfare, the government also has the objectives of maximizing economic growth, optimizing social equity, and multi-objective optimization. The corresponding mathematical objective functions are shown in the Appendix~\ref{other_gov_tasks}.

\begin{figure*}[t]
    \centering
    \includegraphics[scale=0.47]{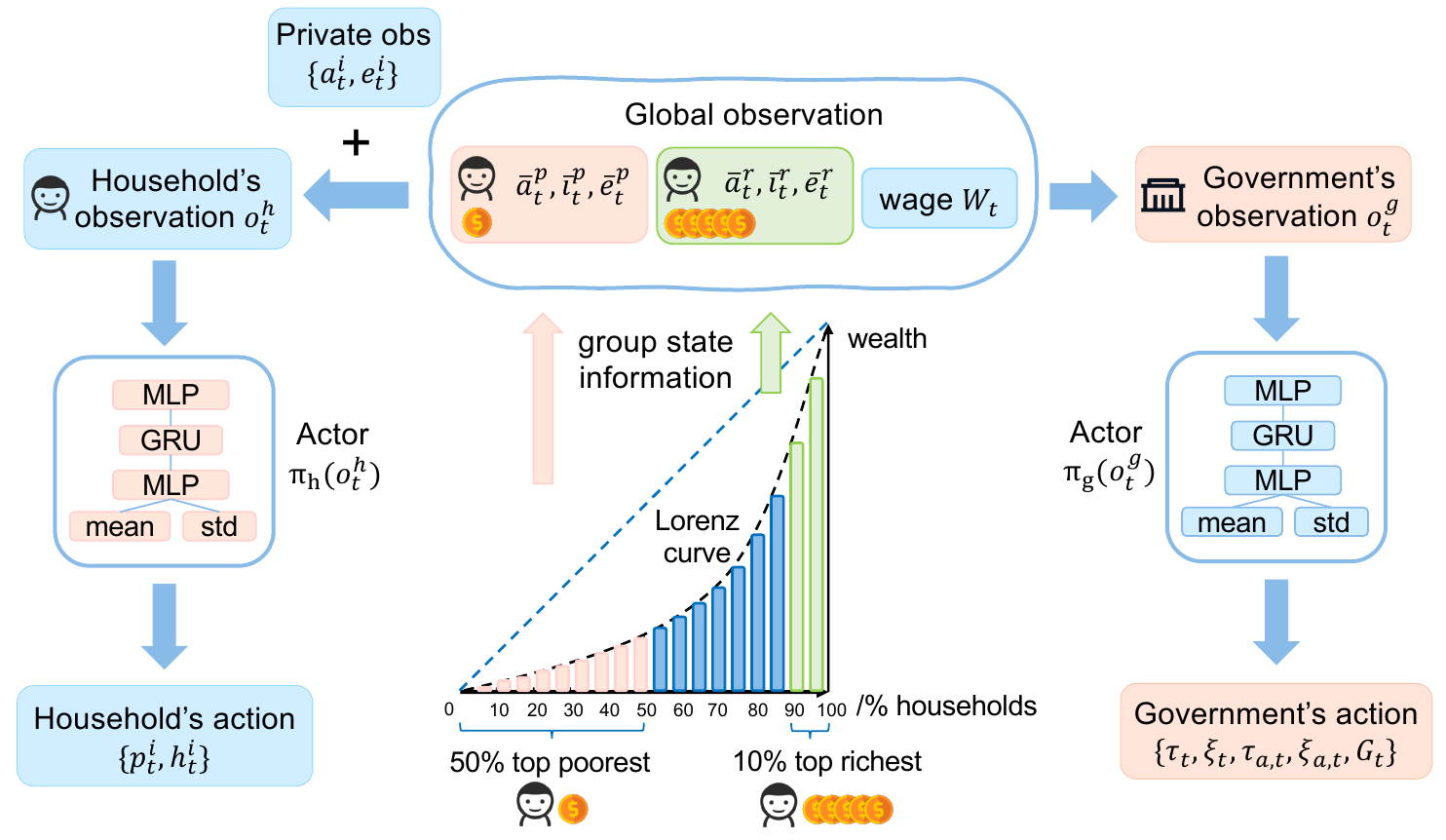}
    \caption{The Markov game between the government and household agents. In the center of the figure, we display the Lorenz curves of households' wealth distribution.  
    The global observation consists of the average assets $\bar{a}_t$, income $\bar{i}_t$, and productivity level $\bar{e}_t$ of the 50\% poorest households and 10\% richest households, along with the wage rate $W_t$. For the government agent, it observes the global observation and takes tax and spending actions $\{\tau_t, \xi_t, \tau_{a,t}, \xi_{a,t}, r^G_t\}$ through the actor-network. For household agents, they observe both global and private observation, including personal assets $\{a^i_t\}$ and productivity level $\{e^i_t\}$, and generate savings and workings actions $\{p^i_t, h^i_t\}$ through the actor-network. The actor-network structure in the figure is just an example.}
    \Description{The Markov game between the government and household agents. In the center of the figure, we display the Lorenz curves of households' wealth distribution.  
    The global observation consists of the average assets $\bar{a}_t$, income $\bar{i}_t$, and productivity level $\bar{e}_t$ of the 50\% poorest households and 10\% richest households, along with the wage rate $W_t$. For the government agent, it observes the global observation and takes tax and spending actions $\{\tau_t, \xi_t, \tau_{a,t}, \xi_{a,t}, r^G_t\}$ through the actor-network. For household agents, they observe both global and private observation, including personal assets $\{a^i_t\}$ and productivity level $\{e^i_t\}$, and generate savings and workings actions $\{p^i_t, h^i_t\}$ through the actor-network. The actor-network structure in the figure is just an example.}
    \label{fig:markov game}
\end{figure*}
\subsection{Financial Intermediary}
We posit a financial intermediary where households can deposit their savings and the intermediary uses these funds to purchase capital and government bonds. Its budget constraint is defined as:
\begin{equation}
    K_{t+1}+B_{t+1}-A_{t+1}=\left(R_t+1-\delta\right) K_t+\left(1+r_{t-1}\right)\left(B_t-A_t\right)
\end{equation}
where $A_t$ are the gross deposits from the households. No-arbitrage implies that $R_{t+1}=r_t+\delta$.

\section{TaxAI Simulator}

In this section, we model the above problem of optimizing tax policies for the government and developing saving and working strategies for households as multiplayer general-sum Partially Observable Markov Games (POMGs). 
In the POMGs $\left\langle \mathcal{N}, \mathcal{S}, \mathcal{O}, \mathcal{A}, P, \mathcal{R}, \gamma\right\rangle$. $\mathcal{N} = \{2,...,N\}$ denotes the set of all agents, $\mathcal{S}$ represents the state space, $\mathcal{O}^i$ denotes the observation space for agent $i$, and $\mathcal{O} := \mathcal{O}^1 \times ... \times \mathcal{O}^\mathbf{N}$.
$\mathcal{A}^i$ signifies the action space for agent $i$, and $\mathcal{A} := \mathcal{A}^1 \times ... \times \mathcal{A}^\mathbf{N}$.
$P: \mathcal{S} \times \mathcal{A} \rightarrow \Omega(\mathcal{S})$ denotes the transition probability from state $s \in \mathcal{S}$ to next state $s' \in \mathcal{S}$ for any joint action $a \in \mathcal{A}$ over the state space $\Omega(\mathcal{S})$. The reward function $\mathcal{R} := \{R^i\}_{i \in \mathcal{N}}$, here $R^i: \mathcal{S} \times \mathcal{A} \times \mathcal{S} \rightarrow \mathbb{R}$ denotes the reward function of the agent $i$ for a transition from $(s, {a})$ to $s^{\prime}$. The discount factor $\gamma \in [0,1)$ keeps constant across time. The specific details of POMGs is shown in the Figure~\ref{fig:markov game} and following paragraphs.

\paragraph{Observation Space $\mathcal{O}$}
In the real world, households can observe their own asset $a_t$ and productivity ability $e_t$, and acquire statistical data about the population from the news. While the government can collect data from all households and access current market prices $W_t$. However, the presence of a large number of heterogeneous households results in a considerably high-dimensional state space. To mitigate the dimensionality challenge, TaxAI categorizes households based on their wealth into two groups~\cite{chancel2022world}: the top 10\% richest and the bottom 50\% poorest households. The average wealth $\{\bar{a}^{r}_t, \bar{a}^{p}_t\}$, income $\{\bar{i}^{r}_t, \bar{i}^{p}_t\}$, and labor productivity levels $\{\bar{e}^{r}_t, \bar{e}^{p}_t\}$ of these two groups are incorporated into the global observation.
Therefore, the government's observation space $\mathcal{O}_g = \{W_t, \bar{a}^{r}_t, \bar{i}^{r}_t,\bar{e}^{r}_t, \bar{a}^{p}_t, \bar{i}^{p}_t,\bar{e}^{p}_t\}$, while the household agent $i$ can observe the global and its private information $\mathcal{O}_h^i = \{W_t, \bar{a}^{r}_t, \bar{i}^{r}_t,\bar{e}^{r}_t, \bar{a}^{p}_t, \bar{i}^{p}_t,\bar{e}^{p}_t, a_t^i, e_t^i\}, i \in \{1,...,N\}$. Moreover, the initialization of state information is calibrated by statistical data from the 2022 Survey of Consumer Finances (SCF)~\footnote{\href{https://www.federalreserve.gov/econres/scfindex.htm}{https://www.federalreserve.gov/econres/scfindex.htm}}.

\paragraph{Action Space $\mathcal{A}$}
The decision-making of household and government agents needs to adhere to budget constraints at every step; however, the abundance of constraints often renders many MARL algorithms ineffective. Therefore, we have introduced proportional actions to alleviate these constraints. For household agents, the optimization of savings $a_{t+1}$ and consumption $c_t$ has been transformed into optimizing savings ratio $p_t \in (0,1)$ and working time $h_t \in [0,1] \cdot h_{max}$, where $h_{max}$ is calibrated by the real wealth-to-income ratio data. 
\begin{equation}
p_t = \frac{a_{t+1}}{i_t-T(i_t)+a_t-T^a(a_t)}
\end{equation}
For the government agent, the fiscal policy tools include optimizing tax parameters $\{\tau_t, \xi_t, \tau_{a,t}, \xi_{a,t}\}$, and the ratio of government spending to GDP $r^G_t = G_t/Y_t$.
Thus, the action space of the government $\mathcal{A}_g$ is $\{\tau_t, \xi_t, \tau_{a,t}, \xi_{a,t}, r^G_t\}$, while the action space of each household $\mathcal{A}_h^i$ is $\{p^i_t, h^i_t\}$.

\paragraph{Reward function $\mathcal{R}$}
The reward function for each household is denoted as:
\begin{equation}
     r_{h,t}(s_t,a^i_{h,t}) = \frac{{c_t^i}^{1-\theta}}{1-\theta}-\frac{{h_t^i}^{1+\gamma}}{1+\gamma}
\end{equation}
On the other hand, the government's objectives are more diverse, and we have defined four distinct experimental tasks within TaxAI: 
(1) Maximizing GDP growth rate.
(2) Minimizing social inequality.
(3) Maximizing social welfare.
(4) Optimizing multiple tasks.
For more details about reward function see Appendix~\ref{other_gov_tasks}.
For example, the government's reward function when maximizing GDP growth rate is denoted as:
\begin{equation}
     r_{g,t}(s_t,a_{g,t}) = \frac{Y_t - Y_{t-1}}{Y_{t-1}}
\end{equation}


In summary, we outline three key improvements made in constructing the \textbf{TaxAI} simulator: (1) To bridge the gap between economic models and the real world, we opt to calibrate TaxAI using 2022 SCF data.
(2) To mitigate the curse of dimensionality associated with high-dimensional state information, we draw inspiration from the World Inequality Report 2022~\cite{chancel2022world} and employ grouped statistical averages for households as a representation of this high-dimensional state information.
(3) In response to the abundance of constraints, we introduce the concept of proportional actions, facilitating control over the range of actions to adhere to these constraints.
More details about environment setting are shown in Appendix~\ref{env setting}, including model assumptions, terminal conditions, parameters setting, and timing tests in Appendix~\ref{timing}.

\vspace{-0.5em}
\section{Experiments}
This section will begin by introducing \textbf{9} baseline algorithms~\ref{baselines}, followed by conducting the following three sub-experiments: 
\textbf{Firstly}~\ref{benchmark_alg}, we aim to illustrate the superior performance of MARL algorithms over traditional methods from both macroeconomic and microeconomic perspectives.
In the \textbf{second} part~\ref{economics analysis}, we conduct an economic analysis of the optimization process for government and heterogeneous household strategies.
\textbf{Lastly}~\ref{scalability}, we assess the scalability of TaxAI by comparing the simulation results for different numbers of households, specifically N=10, 100, 1000, and even 10000. Additional results on full training curves, economic evolution, and hyperparameters, are shown in Appendix~\ref{additional results}.

\vspace{-0.5em}
\subsection{Baselines}\label{baselines}
We compare 9 different baselines, including traditional economic methods and 4 distinct MARL algorithms, providing a comprehensive MARL benchmark for large-scale heterogeneous multi-agent dynamic games in a tax revenue context. Additional experimental settings are shown in the Appendix~\ref{baseline setting}.

{(1) Traditional Economic Methods:} Free Market Policy, Genetic Algorithm (GA)~\cite{dyrda2016optimal}.

{(2) Independent Learning:} Independent PPO~\cite{zhengAIEconomistTaxation2022}.

{(3) Centralized Training Distributed Execution:} MADDPG~\cite{loweMultiAgentActorCriticMixed2017}, MAPPO~\cite{yu2022surprising}, both with parameter sharing.

{(4) Heterogeneous-Agent Reinforcement Learning:} HAPPO, HATRPO, HAA2C~\cite{zhong2023heterogeneous}.

{(5) Mean Field Multi-Agent Reinforcement Learning:} Bi-level Mean Field Actor-Critic (BMFAC), shown in Appendix~\ref{BMFAC}.


\subsection{Comparative Analysis of Multiple Baselines}\label{benchmark_alg}

\begin{figure*}[t]
    \centering
     \includegraphics[scale=0.4]{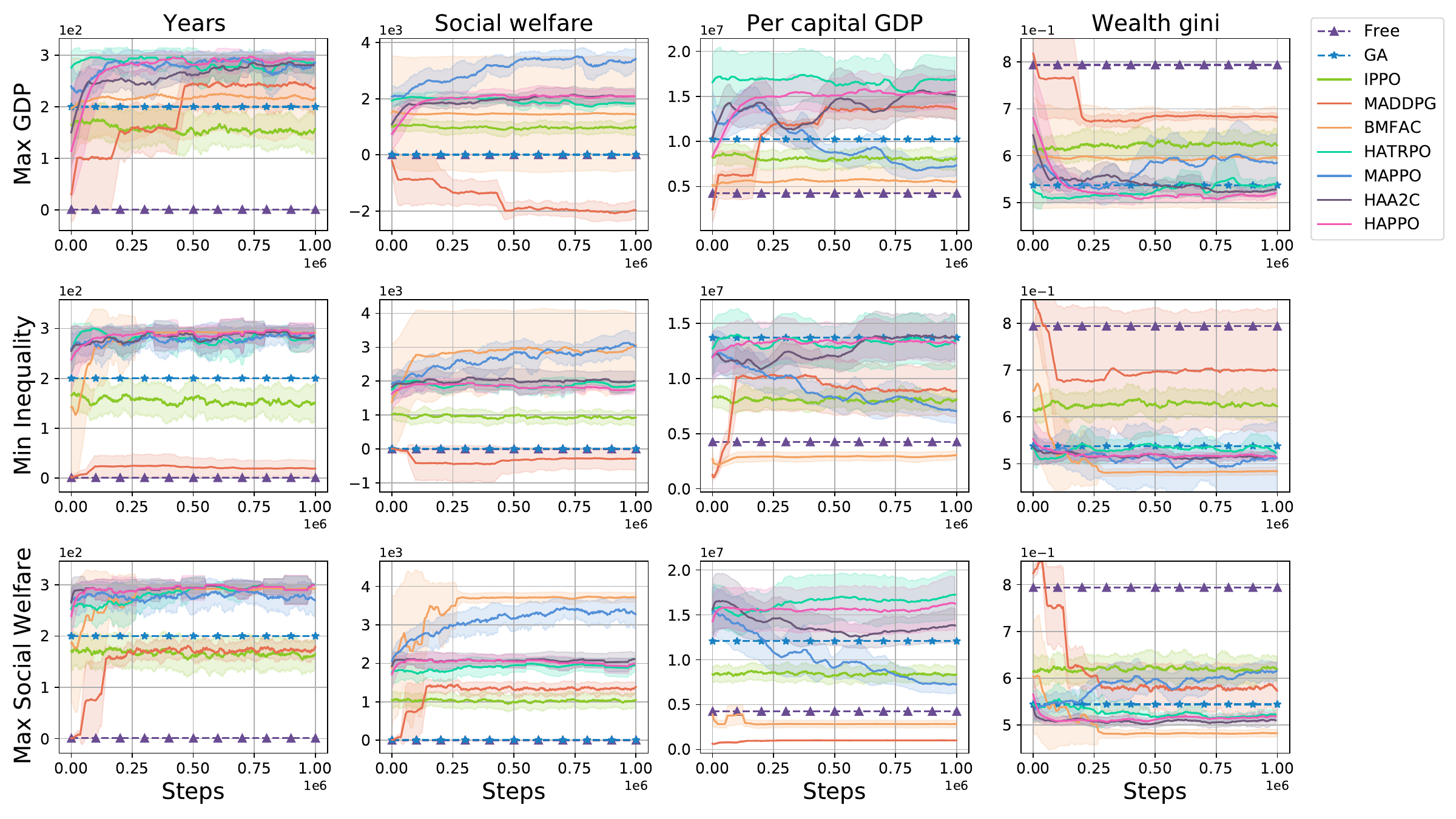}
    \caption{The training curves for 9 baselines on 4 \textbf{macro-economic indicators} under 3 different tasks ($N=100$).}
    \label{fig:small_result}
\end{figure*}
\begin{figure*}
    \centering
     \includegraphics[scale=0.4]{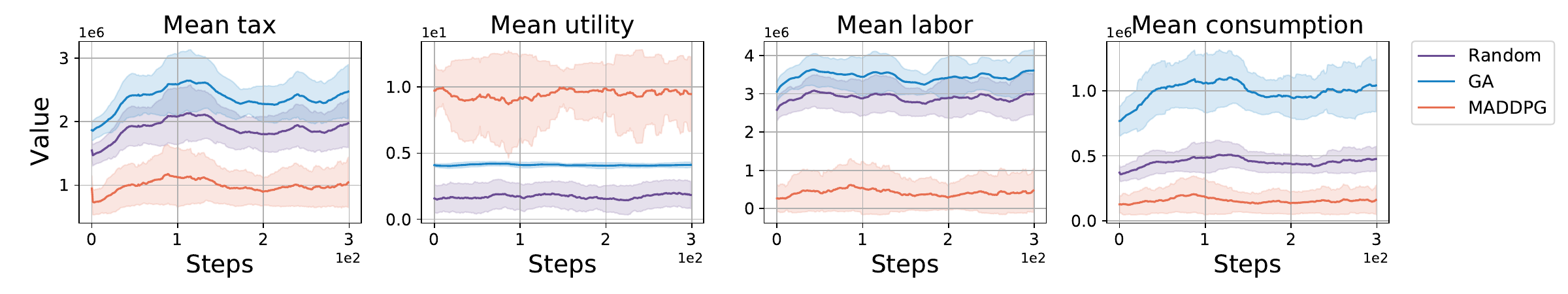}
    \caption{The\textbf{ micro-level behaviors} of Random, GA, and MADDPG households while facing identical observations in an episode (300 steps). The subfigure illustrates the average values of labor provided, consumption, taxes paid, and utility for all households at each step. The results reveal that MADDPG households exhibit \textbf{tax evasion behavior} and attain the \textbf{highest utility}.}
    \label{fig:small_micro_result}
\end{figure*}

We benchmark 9 baselines on 4 distinct tasks, with the training curves of macroeconomic indicators in 3 tasks shown in Figure~\ref{fig:small_result} and test results shown in Table~\ref{table: test table} (Figure~\ref{fig:result} in Appendix~\ref{additional results} presents the training curves for 6 macro-indicators in 4 tasks).
In Figure~\ref{fig:small_result}, each row represents a task, including maximizing GDP, minimizing inequality, and maximizing social welfare. Each column represents a macroeconomic indicator, where longer years indicate longer economic stability, higher GDP represents a higher level of economic development, and a lower wealth Gini coefficient indicates fairer wealth distribution. The X-axis of each subplot represents training steps.
\vspace{-0.5em}
\paragraph{Macroeconomic Perspectives}
From Figure~\ref{fig:small_result}, it can be observed that in each macro-indicator, most MARL algorithms outperform traditional economic methods. In the GDP optimization task, HATRPO achieves the highest per capita GDP, while BMFAC performs best in the tasks of minimizing inequality and maximizing social welfare. Different algorithms also differ in terms of convergence solutions. MADDPG excels in optimizing GDP but at the cost of reducing social welfare for higher GDP. The BMFAC algorithm excels in optimizing social welfare and the Gini coefficient. HARL algorithms, including HAPPO, HATRPO, and HAA2C, can simultaneously optimize all four macroeconomic indicators, but while achieving the highest GDP, social welfare is not maximized. On the other hand, MAPPO excels in optimizing social welfare.

\vspace{-0.5em}
\paragraph{Microeconomic Perspectives}
During the testing phase, we conduct experiments on households following random, GA, and MADDPG policies within the same environment at each step. We utilize 10 distinct random seeds to simulate an economic society spanning 300 timesteps. In Figure~\ref{fig:small_micro_result}, these subplots present various microeconomic indicators, including the average tax revenue, average utility, average labor supply, and average consumption for all households at each time step.
The random policy represents a strategy unaffected by government tax policies, while the GA policy represents a conventional economics approach. We observe that households under the MADDPG strategy pay the lowest taxes, indicating \textbf{tax evasion behavior}, while simultaneously achieving utility levels significantly surpassing those of the GA and random policies. Labor supply and consumption are statistical measures of household microbehavior. We find that MADDPG-based households tend to opt for low consumption and reduced labor supply strategies.

\begin{table}[h]
\caption{Test results for 9 baselines on 5 economic indicators under  {maximizing social welfare} task ($N=100$ households).}
\label{table: test table}
\begin{center}
\begin{small}
\begin{tabular}{c c c ccc}
\toprule
Baselines &{Years} & {Average}&{Per Capita} &{Wealth}&{Income}\\
 &{} & {Social Welfare}&{GDP} &{Gini}&{Gini}\\
\midrule
{Free market} &$ 1.0 $&$ 2.9 $&$ 4.3e6 $&$ 0.79 $&$\mathbf{0.39}$\\
{GA} &$ 200.0 $&$ 6.9 $&$ 1.2e7 $&$   0.54 $&$ 0.52 $\\
{IPPO}&$ 162.7 $&$ 1035.5 $&$ 8.4e6 $&$0.62 $&$ {0.44} $\\
{MADDPG} &$ 204.2 $&$ 1344.6 $&$ 1.0e6 $&$0.61$&$ 0.58 $\\
{MAPPO} &$ 274.5 $&$ 3334.7 $&$ 7.3e6  $&$ 0.61 $&$ 0.65 $\\
{HAPPO} &$ 298.7 $&$ 1986.0  $&$  1.6e7 $&$  0.52 $&$ 0.54 $\\
{HATRPO} &$ \mathbf{300.0} $&$  1945.0 $&$ \mathbf{1.7e7}  $&$ 0.52 $&$  0.54$\\
{HAA2C} &$ \mathbf{300.0} $&$ 2113.3 $&$ 1.4e7  $&$ 0.51$&$ 0.53 $\\
{BMFAC} &$292.8 $&$ \mathbf{3722.2}  $&$ 2.8e6  $&$ \mathbf{ 0.48} $&$ 0.50 $\\  
\bottomrule
\end{tabular}
\end{small}
\end{center}
\vskip -0.1in
\end{table}

\subsection{Economic Analysis of MARL Policy}\label{economics analysis}
\begin{figure*}
    \centering
    \includegraphics[scale=0.47]{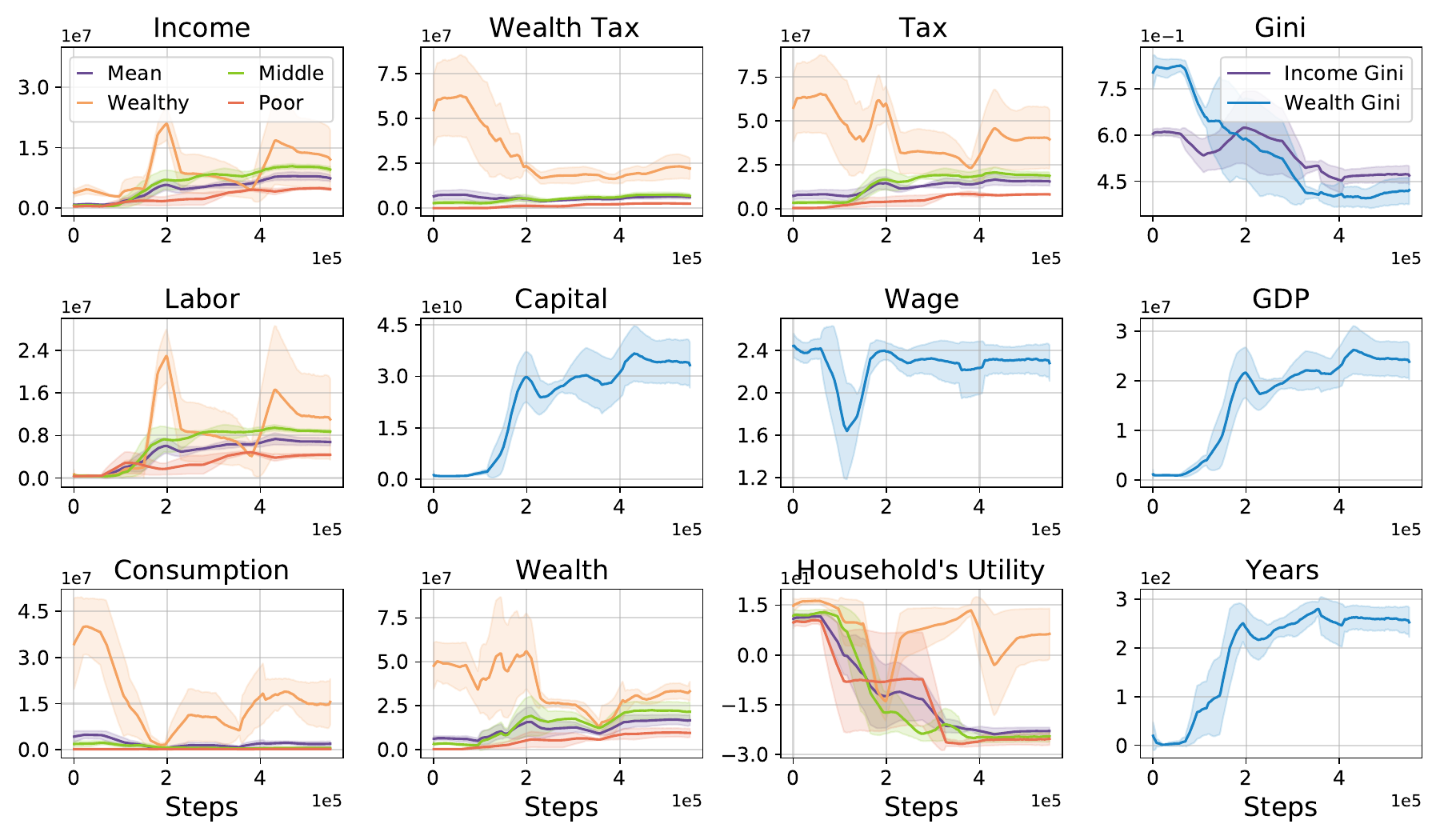}
    \caption{Temporal evolution of economic indicators during MADDPG training under maximizing GDP task on TaxAI ($N=100$).}
    \label{fig: maddpg}
\end{figure*}

\begin{table*}[t]
\footnotesize
\caption{The per capita economic indicators of the MADDPG algorithm during the testing phase at $N=100$ for 3 household groups and the average level across all households.}
\label{table: heterogeneous households}
\centering
\captionsetup[table]{skip=0pt}
\begin{center}
\begin{small}
\begin{tabular}{c c c c c ccccc}
\toprule
Households’ groups  & Income tax & Wealth tax & Total tax & Labor supply  & Consumption & Wealth & Income  & Per year utility \\
\midrule
The wealthy        & $1.9 e6$   & $\mathbf{1.0e7}$    & $\mathbf{1.2e7}$   & $2.3e6$      & $\mathbf{4.4e7}$     & $\mathbf{5.3e7}$ & $5.5 e6$  & $\mathbf{8.7}$              \\
The middle class   & $\mathbf{5.7 e6}$   & $3.0 e6$   & $8.7e6$   & $\mathbf{7.1e6}$      & $6.4e5$     & $2.1e7$ & $\mathbf{7.2 e6}$  & $-24.1$            \\
The poor           & $2.8 e6$   & $1.2e6$    & $4.0e6$   & $4.9e6$      & $2.3e5$     & $9.2e6$ & $4.6 e 6$ & $-25.6 $           \\
Mean value         & $3.8 e6$   & $2.9 e6$   & $6.7e6$   & $5.5e6$      & $4.8e6$     & $1.8e7$ & $5.7 e 6$ & $-22.7$            \\
\bottomrule
\end{tabular}
\end{small}
\end{center}
\end{table*}
Figure~\ref{fig: maddpg} illustrates the training curves of the MADDPG algorithm, aiming to maximize GDP on TaxAI. We utilize the experimental results to analyze the coordination of government actions (tax) and households' actions (labor and consumption) and their impact on economic indicators. The X-axis represents steps, while the Y-axis represents various economic indicators. In the subplots for income tax, wealth tax, total tax, labor, consumption, and households' utility, we categorize households into three groups based on wealth: the wealthy (top 10\% richest), the middle class (top 10-50\% richest), and the poor (50-100\% top richest). We display the average indicators for these three groups as well as the average for all households (purple line), other subplots represent macroeconomic indicators (blue line). In each subplot, the solid line represents the mean of experimental results under different seeds, and the shaded area represents the standard deviation. The results of the testing phase are presented in Table~\ref{table: heterogeneous households}. The following are two intriguing findings:

\paragraph{1. MADDPG converges towards the highest GDP while compromising social welfare under maximizing GDP task} 
As observed from Figure~\ref{fig: maddpg}, during the initial $2e5$ steps, the government increases income and wealth tax, leading to a reduction in the income and wealth Gini, making the wealth distribution more equitable. Social equity is a prerequisite for increasing the duration of the economic system (measured by years). Simultaneously, households choose to increase labor and reduce consumption, leading to a decrease in their utility. As the total savings of households increase, financial intermediaries can provide more production capital to firms. The additional production labor and capital leads to an increase in output (GDP). The wage rate tends to decrease with an increase in labor and increase with a capital increase, exhibiting a trend of initially decreasing and then increasing.
\balance

\paragraph{2. Different wealth groups adopt distinct strategies.} From the experimental curves (Figure~\ref{fig: maddpg}) and results (Table~\ref{table: heterogeneous households}) for the wealthy, the middle class, and poor households, we find the following patterns: During $0 \sim 2e5$ steps, the wealthy contribute significantly to taxation. To stabilize their wealth, they increase work hours and reduce consumption, leading to a decline in utility. In the second phase ($2e5 \sim 4e5$ steps), the wealthy maximize utility by reducing work hours and significantly increasing consumption, even though their wealth levels decrease. In the third phase ($4e5 \sim 6e5$ steps), the wealthy simultaneously increase labor and consumption, resulting in increased wealth while maintaining relatively stable utility. On the other hand, the middle class and the poor slightly increase work hours and reduce consumption during all three phases, leading to modest growth in wealth but significantly lower utility compared to the wealthy.

\begin{table}[h]
\footnotesize
\caption{The per capita GDP achieved by 9 baselines for different numbers $N$ of households under maximizing GDP task.}
\label{table: scalability}
\centering
\captionsetup[table]{}
\begin{center}
\begin{small}
\begin{tabular}{c c c c c}
\toprule
Algorithm & {$N$=10} &  {$N$=100} &  {$N$=1000} &  {$N$=10000} \\ 
\midrule
{Free Market}  &{$1.3e6$}&{$4.3e6$}&{$3.9e6 $}&{$4.0e6 $}\\
{GA}&{$\mathbf{1.7e8 }$}&{$1.5e7 $}&{NA}&{NA} \\
{IPPO}  &{$ 5.0e6 $}&{${1.6e7 }$}&{$1.7e7 $}&{$1.6e7$} \\
{MADDPG}   &{$3.2e6 $}  &{$1.1e7$}&{${1.7e7 }$}&{$\mathbf{1.7e7}$} \\
{MAPPO} & {$6.1e6 $}  &{$7.3e6$}&{${1.2e7 }$}&{NA}  \\
{HAPPO} & {$1.8e7 $}  &{$1.6e7$}&{${1.5e7 }$}&{NA} \\
{HATRPO} & {$3.2e7 $}  &{$\mathbf{1.7e7}$}&{$\mathbf{2.0e7 }$}&{NA} \\
{HAA2C}  & {$1.6e7 $}  &{$1.4e7$}&{${1.6e7 }$}&{NA} \\
{BMFAC}   &{$4.0e6 $}  &{$1.2e7 $}&{$1.2e7 $}&{NA}  \\
\bottomrule
\end{tabular}
\end{small}
\end{center}
\end{table}

\subsection{Scalability of Environment}\label{scalability}
To showcase the scalability of TaxAI in simulating large-scale household agents, we conduct tests with varying numbers of households: 10, 100, 1000, and even 10,000 (as shown in Table~\ref{table: scalability}). The table presents the average per capita GDP for each baseline. The results in Table~\ref{table: scalability} indicate that IPPO and the improved MADDPG algorithm successfully achieve the maximum GDP when $N$ = 10,000, whereas traditional methods yield NA (not available). HATRPO achieves optimal strategies at $N$ = 100 and $N$ = 1000, respectively, while GA only achieves optimal GDP when $N$ is small. The above results indicate that TaxAI is capable of simulating 10,000 household agents, surpassing other benchmarks by a significant margin. Moreover, MARL algorithms can successfully solve the optimal tax problem in large-scale agent scenarios. These two advantages are crucial for simulating real-world society.

\section{Conclusion}
We introduce TaxAI, a large-scale agent-based dynamic economic environment, and benchmark 2 traditional economic methods and 7 MARL algorithms on it. 
TaxAI, in contrast to prior work, excels in modeling large-scale heterogeneous households, a wider range of economic activities, and tax types. Moreover, it is calibrated using real data and comes with open-sourced simulation code and MARL benchmark.
Our results illustrate the feasibility and superiority of MARL in addressing the optimal taxation problem, while also revealing MARL households' tax evasion behavior.

In the future, we aim to expand and enrich the economic theory of TaxAI by incorporating a broader range of social roles and strategies. Furthermore, we will enhance the scalability of our simulator to accommodate one billion agents, enabling simulations that closely resemble real-world scenarios. By doing so, we aim to attract more researchers to explore complex economic problems using AI or RL techniques, thereby offering practical and feasible recommendations for social planners and the population.

\begin{acks}
Haifeng Zhang thanks the support of the
National Natural Science Foundation of China, Grant No. 62206289.
\end{acks}
\section*{Ethics Statement}
This paper presents work whose goal is to advance the fields of AI for economics. Our work aims to offer suggestions and references for governments and the people, yet it must not be rashly applied to the real world. There are many potential societal consequences of our work, none of which we feel must be specifically highlighted here.






\bibliographystyle{ACM-Reference-Format} 
\bibliography{sample}

\newpage
\appendix

\begin{table*}[t]
\footnotesize
\caption{Comparison of 5 classical tax models on key economic indicators.}
\label{table: tax_models}
\centering
\captionsetup[table]{}
\begin{center}
\begin{small}
\begin{tabular}{c c c c c}
\toprule
Classic Tax models & Incomplete Markets & Uncertainty & Inequality & Emphasis \\
\midrule
Ramsey–Cass–Koopmans (RCK) model &  {$\times$}   &  {$\times$}   &  {$\times$}   & Economic Growth\\
Mirrlees model / Diamond-Mirrlees model & {$\times$} & {\checkmark}& {\checkmark}& Tradeoff between Economic Growth and Equity\\
Overlapping Generations (OLG) model& {\checkmark}& {\checkmark} &{\checkmark}& Intergenerational Equity, Pension Systems\\
Bewley-Aiyagari model &{\checkmark} &{\checkmark} &{\checkmark} &Tax Policies, Social Equity, Social Security Systems\\
Atkinson-Stiglitz model&  {$\times$}  & {$\times$}  &{\checkmark} &Social Welfare and Taxation Systems\\
\bottomrule
\end{tabular}
\end{small}
\end{center}
\end{table*}

\section{Environment Setting}\label{env setting}
\subsection{Classical Tax Models}
In Section~\ref{related_work}, we compare classical tax models using key economic indicators (shown in Table~\ref{table: tax_models}) to highlight the Bewley-Aiyagari model's advantages in modeling economic activities and determining optimal tax strategies.

\begin{itemize}
    \item Incomplete Markets: Markets with asymmetric information.
    \item Uncertainty: Risks and unpredictability in decision-making and forecasting.
    \item Inequality: Disparities in wealth and income distribution.
\end{itemize}

\subsection{Government Tasks}\label{other_gov_tasks}
The mathematical objective functions for the four types of government tasks are as follows:
\paragraph{Maximizing GDP Growth Rate} The economic growth can be measured by Gross Domestic Product (GDP). Without considering imports and exports in an open economy, GDP is equal to the output $Y_t$ in our model. 
Based on reality, we set the government's objective to maximize the GDP growth rate,
\begin{equation}
    \max \quad J = \mathbb{E}_0 \sum_{t=0}^{T_N} \beta^t \left(\frac{Y_t - Y_{t-1}}{Y_{t-1}}\right)
\end{equation}
\paragraph{Minimizing Social Inequality} Social equality and stability build the foundation for all social activities. Social inequality is usually measured by the Gini coefficient of wealth distribution $\mathcal{W}_t$ and income distribution $\mathcal{I}_t$. The Gini coefficient is calculated by the ratio of the area between the Lorenz curve and the perfect equality line, divided by the total area under the perfect equality line (shown in figure~\ref{fig:markov game}). The Gini coefficient ranges between 0 (perfect equality) and 1 (perfect inequality).
\begin{equation}
    \min \quad J = \mathbb{E}_0 \sum_{t=0}^{T_N} \beta^t \left( \text{Gini}(\mathcal{I}_t) \text{Gini}(\mathcal{W}_t)\right)
\end{equation}
\paragraph{Maximizing social welfare} Optimizing social welfare refers to maximizing the sum of utilities for all households. This is one of the government's most important responsibilities.
\begin{equation}
    \max  \quad J = \mathbb{E}_0 \sum_{t=0}^{T_N} \beta^t \left(\sum_i^N \frac{{c_t^i}^{1-\theta}}{1-\theta}-\frac{{h_t^i}^{1+\gamma}}{1+\gamma} \right)
\end{equation}
\paragraph{Optimizing Multiple Tasks} If the government aims to simultaneously optimize multiple objectives, we weigh and sum up multiple objectives. The weights $\omega_1$, and $\omega_2$ indicate the relative importance of Gini and welfare objectives.
\begin{equation}
\small
\begin{aligned}
    \max \quad J = \mathbb{E}_0 \sum_{t=0}^{T_N} \beta^t \Bigg(\frac{Y_t - Y_{t-1}}{Y_t} - \omega_1 \cdot \text{Gini}(\mathcal{I}_t) \text{Gini}(\mathcal{W}_t) \\
    +  \omega_2 \cdot \left(\sum_i^N \frac{{c_t^i}^{1-\theta}}{1-\theta}-\frac{{h_t^i}^{1+\gamma}}{1+\gamma} \right) \Bigg)
\end{aligned}
\end{equation}

In TaxAI, the four types of government reward functions are as follows:

{(1) Maximizing GDP Growth Rate:}
\begin{equation}
     r_{g,t}(s_t,a_{g,t}) = \frac{Y_t - Y_{t-1}}{Y_{t-1}}
\end{equation}

{(2) Minimizing Social Inequality:} 
\begin{equation}
     r_{g,t}(s_t,a_{g,t}) = - \text{Gini}(\mathcal{I}_t) \text{Gini}(\mathcal{W}_t)
\end{equation}

{(3) Maximizing Social Welfare:}
\begin{equation}
     r_{g,t}(s_t,a_{g,t}) = \sum_i^N \frac{{c_t^i}^{1-\theta}}{1-\theta}-\frac{{h_t^i}^{1+\gamma}}{1+\gamma}
\end{equation}

{(4) Optimizing Multiple Tasks:}
\begin{equation}
\begin{aligned}
    r_{g,t}(s_t,a_{g,t}) = \frac{Y_t - Y_{t-1}}{Y_t} - \omega_1 \cdot \text{Gini}(\mathcal{I}_t) \text{Gini}(\mathcal{W}_t) \\
    +  \omega_2 \cdot \left(\sum_i^N \frac{{c_t^i}^{1-\theta}}{1-\theta}-\frac{{h_t^i}^{1+\gamma}}{1+\gamma} \right)
\end{aligned}
\end{equation}

\subsection{Assumption}\label{assumption}
We have summarized the assumptions involved in the economic model section~\ref{BA model} as follows:
\begin{enumerate}
    \item Social roles, such as households and government, are considered rational agents.
    \item Households are not allowed to incur debt and engage only in risk-free investments.
    \item The labor productivity of households is categorized into two states: normal state and super-star state. The dynamic changes in each state follow an AR(1) process.
    \item The capital market, goods market, and labor market clear.
    \item The technology firm represents all enterprises and factories, producing a homogeneous good, and follows the Cobb-Douglas production function.
    \item The financial Intermediary operates without arbitrage.
\end{enumerate}

\subsection{Model Dynamics}\label{dynamics}
\begin{figure*}[h]
    \centering
    \includegraphics[scale=0.54]{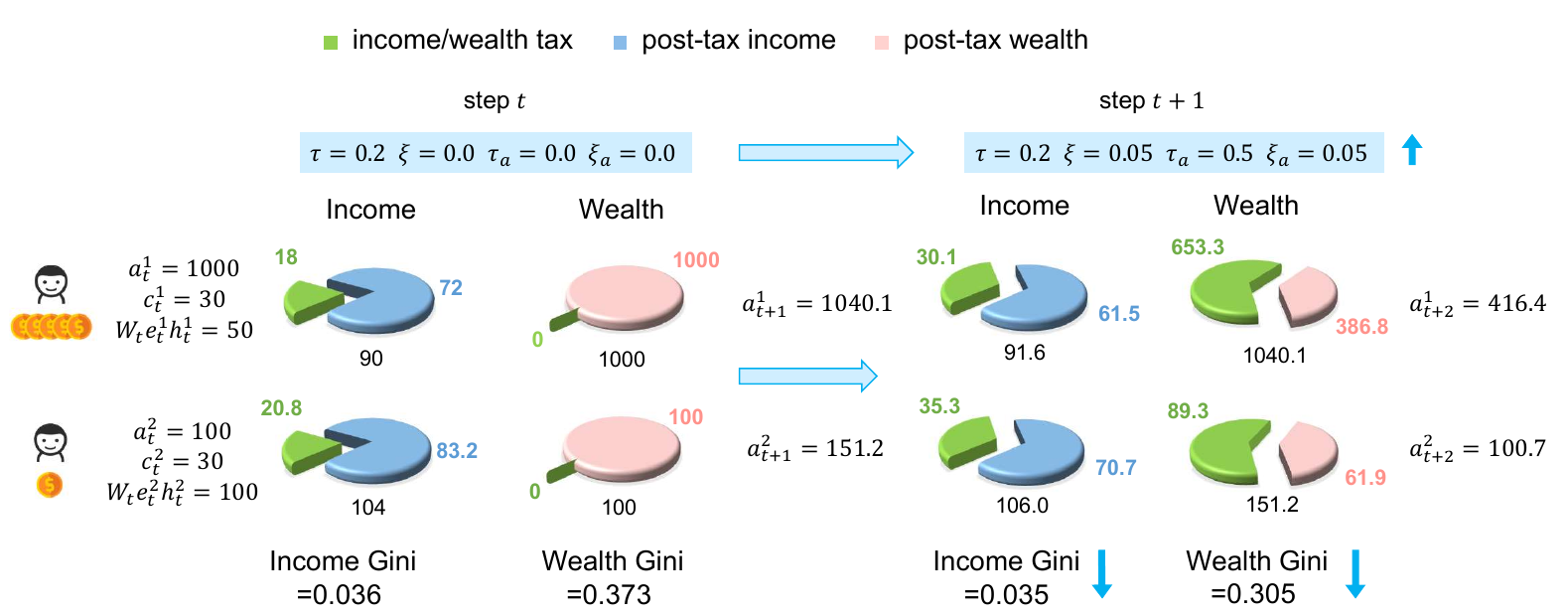}
    \caption{Illustration of tax impact on the rich and poor households.}
    \label{fig:old_dynamics}
\end{figure*}
To elucidate the dynamics of the economic model more lucidly, we present in Figure~\ref{fig:old_dynamics} the impact of tax changes (from time $t$ to time $t+1$) on individuals with varying levels of wealth. The pie chart in the figure illustrates the current households' income and wealth distributions, along with the taxation proportions. The green sector represents income or wealth taxation, the blue sector signifies post-tax income, and the pink sector represents post-tax wealth. Furthermore, we calculate the Gini coefficients for income and wealth distributions to illustrate the effects of changes in tax policies on social equity. (Given that this economic model involves only two households, the Gini coefficients are much smaller than those found in real-world scenarios; therefore, their use here is only for illustrating Gini coefficient variations). 
Suppose the economy consists of two categories of households—a wealthy household (with initial wealth $a_t^1=1000$) and a poor household (with initial wealth $a_t^2=100$). Their income predominantly originates from labor income $W_t e_t h_t$ and asset returns $r_{t-1} a_t$ (where $r_{t-1}=0.04$). And they have the same consumption level of $c^1_t=c^2_t=30$ for each step.

At \textbf{step $t$}, the wealthy household's income is $i_t^1=50 + 0.04*1000=90$, while the poor household's income is $i_t^2=100 + 0.04*100=104$. Considering tax rate parameters of $\tau=0.2$ and $\xi=\tau_a=\xi_a=0$, both the wealthy and poor households face an income tax rate of $20\%$, with no taxation on assets. Consequently, the income tax for the wealthy household amounts to $90 *0.2=18$, resulting in a post-tax income of $90-18=72$. Similarly, the income tax for the poor household is $20.8$, resulting in a post-tax income of $83.2$. Through the budget constraint in (\ref{households_utility}), we determine the next-step wealth as $a_{t+1}^1=1040.1$ and $a_{t+1}^2=151.2$. The Gini coefficient for income is 0.036, while the Gini coefficient for wealth is $0.373$, reflecting the wealth distribution between these two economic agents.

Moving to \textbf{step $t+1$}, despite maintaining unchanged consumption levels and labor incomes for both wealthy and poor households, the increase in wealth fosters an increment in asset returns. Consequently, the incomes for the wealthy and poor households have risen relative to step $t$; namely, $i_{t+1}^1=91.6$ and $i_{t+1}^2=106.0$. At this juncture, while the tax rate parameter $\tau$ remains unchanged, $\xi$ has been elevated to $0.05$, transforming the income tax function into a non-linear structure and raising tax rates. Meanwhile, the asset tax parameters $\tau_a=0.5$ and $\xi_a=0.05$ prompt the government to initiate asset taxation. Under equivalent labor income conditions, the wealthy household's income tax has risen to $30.1$ via Equation~\ref{tax function}, while asset taxation has surged to $653.3$. A similar scenario ensues for the poor household. Thus, due to the escalated tax rate, more than half of the asset for both wealthy and poor households is allocated toward taxation. Ultimately, the remaining wealth is $a_{t+2}^1=416.4$ and $a_{t+2}^2=100.7$. While initially, the wealthy household's asset was \textbf{tenfold} that of the poor household, the introduction of wealth taxation by the government has reduced this disparity to roughly a \textbf{fourfold} difference. The increase in $\xi$ has led to a marginal decline in the income Gini coefficient, while the wealth tax has significantly lowered the wealth Gini coefficient, concurrently promoting social equity.

\begin{equation}
    \label{tax function}
    T\left(i_t\right)=i_t-(1-\tau) \frac{i_t^{1-\xi}}{1-\xi},  \quad T^a\left(a_t\right)=a_t-\frac{1-\tau_a}{1-\xi_a} a_t^{1-\xi_a}
\end{equation}

\subsection{Environment Parameters}
The assigned parameters of TaxAI are shown in table~\ref{table:assigned parameter}. At the beginning of each episode, we initialize households' assets $\{a^i_0\}^N_i$ via the statistical data from the 2022 Survey of Consumer Finances (SCF) data and calibrate aggregate labors $L_t$ and the maximum of working hours $h_{max}$ based on the wealth-to-income ratio computed from the 2022 SCF data.
\begin{table}[h]
\caption{Assigned Parameters in Environment}
\label{table:assigned parameter}
\begin{center}
\begin{small}
\begin{tabular}{l r}
\toprule
Hyperparameter & {Value} \\  
\midrule
{CRRA $\theta$}  & 1 \\
{Inverse Frisch elasticity $\gamma$}  & 2\\
{discount factor $\beta$}  & 0.975 \\
{capital elasticity $\alpha$}  & $1/3$ \\
{depreciation rate $\delta$}  &  $0.06$  \\
{return to savings $r_t$}  & 0.04  \\
{consumption tax rate $\tau_s$} & 0.065 \\ 
{normal to super-star probability $p$} & $2.2e^{-6}$ \\
{remain in super-star probability $q$} & 0.990 \\ 
{persistence $\rho_e$}  & 0.982 \\
{volatility of the standard normal shocks $\sigma_e$} & 0.200 \\
{super ability times $\bar{e}$} & 504.3 \\
{Wealth to income ratio} & 6.6\\
\bottomrule
\end{tabular}
\end{small}
\end{center}
\end{table}

\subsection{Teriminal Condition}
We will terminate the current episode under the following circumstances:
\begin{itemize}
    \item If the maximum number of steps in an episode is reached.
    \item If the Gini coefficient of income or wealth exceeds a threshold;
    \item If the Gross Domestic Product (GDP) is insufficient to cover the total consumption of households ($C_t + G_t > Y_t$).
    \item In the event of household bankruptcy ($a_t^i < 0$).
    \item If the calculation of households' or government's rewards results in an overflow or leads to the appearance of NaN (Not-a-Number) values. 
\end{itemize}

\section{Timing Tests}\label{timing}
Table~\ref{table: timing} shows the number of dynamics evaluations per second in a single thread.
Results are averaged over 10 episodes on AMD EPYC 7742 64-Core Processor with GPU GA100 [GRID A100 PCIe 40GB].
\begin{table}[h]
\caption{The number of dynamics evaluations per second in a single thread.}
\label{table: timing}
\centering
\captionsetup[table]{skip=0pt}
\begin{center}
\begin{small}
\begin{tabular}{c c c c c}
\toprule
Algorithm & {$N$=10} &  {$N$=100} &  {$N$=1000} &  {$N$=10000} \\  
\midrule
{Number of steps}  &{$1799.927$}&{$482.335$}&{$46.983$}&{$4.018$}\\
{num of episodes}  &{$6.289$}&{$1.689$}&{$0.157$}&{$0.013$}\\
{episode length}  &{$285.900$}&{$285.400$}&{$300.000$}&{$300.000$}\\
\bottomrule
\end{tabular}
\end{small}
\end{center}
\end{table}
\section{Genetic Algorithm}\label{GA}
In this subsection, we provide additional details regarding the implementation of genetic algorithms. To evaluate the performance of different policies, we employ relevant economic indicators such as GDP and the Gini coefficient as the fitness function. Our genetic algorithm implementation includes the utilization of uniform crossover and simple mutation. The specific parameters used in the algorithm are presented in Table~\ref{table: ga hyperparameter}.

For household agents, we refer to Heathcote's~\cite{heathcote2014consumption} approach in selecting saving and working hours for households under the shocks of agent productivity and preferences. However, our model does not incorporate the preference for working and consumption, denoted as $\varphi_t=0$. The productivity of work is determined by
 \begin{equation} \label{productivity}
\log e_t=\alpha_t+\varepsilon_t
\end{equation}
\begin{equation}
    \alpha_t=\alpha_{t-1}+\omega_t
\end{equation}
\begin{equation}
    \varepsilon_t = \kappa_t+\theta_t
\end{equation}
The households strategies are obtained by competitive equilibrium under the productivity shock. The proportions of consumption to income and working hours $c_t^a$ and $h_t^a$ are calculated using the following equations:
\begin{equation} \label{ga action c}
\log c_t^a\left(s^t\right) =-\varphi_t+\frac{1+\gamma}{\gamma+\theta} \alpha_t+\mathcal{M}_t^a
\end{equation}
\begin{equation}\label{ga action h}
\log h_t^a\left(s^t\right)  =-\varphi_t+\frac{1-\theta}{\gamma+\theta} \alpha_t+\frac{1}{\gamma}\left(\kappa_t+\theta_t\right)-\frac{\theta}{\gamma} \mathcal{M}_t^a
\end{equation}
\begin{equation} \label{mt}
\mathcal{M}_t^a=\gamma /(\gamma+\theta) \log \left(\int \exp \left(\frac{1+\gamma}{\gamma} \theta_t\right) d F_{\theta t}\right)
\end{equation}
where $\omega_t$ and $\theta_t$ is an independently distributed shock, 
 $\kappa_t$ is a permanent shock ($\kappa_t=0$ in our experiment).
 
\begin{table}[t]
\caption{Variables in Economic Model}
\label{table:model parameter}
\begin{center}
\begin{small}
\begin{tabular}{l l}
\toprule
Variable & {Meaning} \\  
\midrule
$W_t$ & Wage rate \\
 $e_t$ & Individual labor productivity levels \\
 $h_t$ &  Working hours \\
  $a_t$ & Household's asset(wealth) \\
  $a_0$ & Household's initial asset(wealth) \\
  $r_t$ & interest rate \\
  $i_t$ & Household's income \\ 
  $N$ & Households' number \\
  $\rho_e$ & Persistence \\
  $u_t$ & Standard normal shocks \\
  $\sigma_e$ &  Volatility of the standard normal shocks \\
  $\bar{e}$ & Labor market ability of super-star state \\
  $p$ & Transition probability from normal state\\ 
  {} & to super-star state\\
  $q$ & Transition probability of remaining \\
  {} &  in super-star state\\
   $\beta$ &  Discount factor\\
   $\theta$ &  Coefficient of relative risk aversion (CRRA)\\
   $\gamma$ & Inverse Frisch elasticity\\
   $\tau_s$ & Consumption taxes \\
   $T(i_t)$ & Income taxes \\
   $T^a(a_t)$ & Asset taxes  \\
   $\tau, \tau_a$ &  Average level of marginal income \\
   {} & and asset tax \\
   $\xi, \xi_a$ & Slope of marginal income \\
   {} & and asset tax schedule \\
   $Y_t$ & Firm's output\\
   $K_t$ & Capital for production\\
   $L_t$ &  Labor  for production \\
   $\alpha$ & Capital elasticity \\
    $C_t$ & Households' gross consumption \\
    $G_t$ & Government spending \\ 
   $X_t$ & Physical capital investment \\
   $\delta$ & Depreciation rate \\
   $R_t$ &  Borrowing rate of firm \\
   $G_t$ & Government spending\\
   $T_t$ & Households' gross taxation\\
   $B_t$ & Debt\\ 
   $\mathcal{I}_t$ & Households' gross post-tax income \\
    $\mathcal{W}_t$ & Households' gross post-tax assets \\
    $A_t$ & Households' gross deposits \\
    $\omega_1$, $\omega_2$ & Relative weight of social equality and social welfare\\
\bottomrule
\end{tabular}
\end{small}
\end{center}
\end{table}

\section{Bi-level Mean Field Actor-Critic Algorithm}\label{BMFAC}
We propose a Bi-level Mean Field Actor-Critic (BMFAC) algorithm for the Stackelberg game between the government and $N$ households.
In our BMFAC method, the government serves as the leader and the households as equal followers. The objective of the government is to maximize its objective by setting tax rates and government spending while anticipating the households' actions. Conversely, households aim to optimize their lifetime utility by making decisions regarding their saving policies and working hours.
Initially, the government's action is initialized to a free-market policy (without taxes), and we employ an iterative process to optimize the household agents' policies in the inner loop. Given the household agents' policies, we generate sample trajectories for the government agent and update its actor and critic networks accordingly. Subsequently, we sample data and update the networks for the households, considering the government's policy. This iterative process continues until convergence is attained.

\begin{algorithm}[ht]
\caption{BMFAC: Bi-level Mean Field Actor Critic}
\label{alg:ActQ}
\begin{algorithmic}[1]
    \STATE Initialize $Q_g^{\phi}, Q_h^{\phi^{-}}, \pi^{\theta}_g, \pi^{\theta_{-}}_h$, and $\bar{a}_h^i$ for all $i \in\{1, \ldots, N\}$
    \STATE Initialize the government policy as the free market rules $a_g^0 = \{0,0,0,0,0.189\}$.
    \FOR{Outer Loop}
        \FOR{Inner Loop}
            \FOR{each sample step}
                \STATE Given the government policy $\pi_g$, the government agent sample action $a_g \sim \pi_g(o_g)$; the household agent $i$ sample action $a^i_h \sim \pi_h({o_h^i}, a_g)$. \STATE Take the actions and observe $r_g, \mathbf{r_h} = \{r_h^1, ..., r_h^N\}$ and next state $o_g^\prime, \mathbf{o_h} = \{{o_h^1}^\prime, ..., {o_h^N}^\prime\} $.
                \STATE Store $\left\langle o_g, \mathbf{o_h},a_g, \mathbf{a_h},  r_g, \mathbf{r_h}, o^{\prime}_g, \mathbf{o_h}^\prime  \right\rangle$ in replay buffer $\mathscr{D}$.
            \ENDFOR
            \STATE Update the households' actor and critic for $N^h_{\text{update}}$ steps:
            $$  \psi \leftarrow \psi-\lambda_Q \hat{\nabla}_\psi \mathscr{L}(\psi)$$
            $$  \psi^{-} \leftarrow \psi^{-} -\lambda_Q \hat{\nabla}_{\psi^{-}} \mathscr{L}(\psi^{-})$$
            $$  \sigma \leftarrow \sigma-\lambda_\pi \hat{\nabla}_\sigma \mathscr{L}(\sigma)$$
        \ENDFOR
        \FOR{each sample step}
            \STATE Given the households policy $\pi_h$, the government agent sample action $a_g \sim \pi_g(o_g)$; the household agent $i$ sample action $a^i_h \sim \pi_h({o_h^i}, a_g)$.
            \STATE Take the actions and observe $r_g, \mathbf{r_h} = \{r_h^1, ..., r_h^N\}$ and next state $o_g^\prime, \mathbf{o_h} = \{{o_h^1}^\prime, ..., {o_h^N}^\prime\} $.
            \STATE Store $\left\langle o_g, \mathbf{o_h},a_g, \mathbf{a_h},  r_g, \mathbf{r_h}, o^{\prime}_g, \mathbf{o_h}^\prime  \right\rangle$ in replay buffer $\mathscr{D}$.
        \ENDFOR
        \STATE Update the government's actor and critic for $N^g_{\text{update}}$ steps:
            $$  \phi \leftarrow \phi-\lambda_Q \hat{\nabla}_\phi \mathscr{L}(\phi)$$
            $$  \phi^{-} \leftarrow \phi^{-} -\lambda_Q \hat{\nabla}_{\phi^{-}} \mathscr{L}(\phi^{-})$$
            $$  \theta \leftarrow \theta-\lambda_\pi \hat{\nabla}_\theta \mathscr{L}(\theta)$$
    \ENDFOR
\end{algorithmic}
\end{algorithm}


\subsection{Government Agent}
We update the government agent via the actor-critic method. We consider double Q-function $Q_\phi(\mathbf{o}_t,\mathbf{a}_t)$, $Q_{\phi^{-}}(\mathbf{o}_t,\mathbf{a}_t)$, and a policy $\pi_{\theta}(a_{g,t} \mid o_{g,t})$. For example, the value functions can be modeled as expressive neural networks and the policy as a Gaussian with mean and covariance given by neural networks. We will next derive update rules for these parameter vectors.
\paragraph{The critic network} The Q-function parameters can be trained to minimize the soft Bellman residual.
\begin{equation}
    \mathscr{L}\left( \phi \right)= \left(y_g-Q_{\phi}\left(o_g,a_g, \{\pi_h(o_h^i, a_g)\}_i^N\right)\right)^2
\end{equation}
where the target mean field value with the weight $\phi^-$ is computed by
\begin{equation}
    y_g= r_g+ \gamma \mathbb{E}_{a^{\prime}_g \sim \pi_g} Q_{\phi^{-}}(s^{\prime}_g, a^{\prime}_g, \{\pi_h(o_g^\prime, {o_h^i}^\prime, a_g^\prime)\}_i^N)
\end{equation}
Differentiating $\mathscr{L}\left( \phi \right)$ gives
\begin{equation}
    \begin{aligned}
        \nabla_{\phi} \mathscr{L}\left( \phi \right) = \left( y_g -  Q_{\phi}\left(o_g,a_g, \{\pi_h(o_h^i, a_g)\}_i^N\right) \right) \\ \nabla _\phi Q_{\phi}\left(o_g,a_g, \{\pi_h(o_h^i, a_g)\}_i^N\right)
    \end{aligned}
\end{equation}
which enables the gradient-based optimizers for training.
\paragraph{The actor network} The policy network of the government $\pi^\theta_g$ is training by the sampled policy gradient:
\begin{equation}
    \nabla_\theta \mathscr{L}\left( \theta \right) \approx \nabla_\theta \log{\pi_g^\theta(o_g)} Q_{\phi}\left(o_g,a_g, \{\pi_h(o_h^i, a_g)\}_i^N\right)
\end{equation}

\subsection{Household Agents}
We update household agents via the mean field actor-critic method. We consider double Q-function $Q_\psi(\mathbf{o}_t,\mathbf{a}_t)$, $Q_{\psi^{-}}(\mathbf{o}_t,\mathbf{a}_t)$ as a shared critic and a shared policy $\pi_{\sigma}(a_{h,t} \mid o_{h,t})$ for all household agents. For example, the value functions can be modeled as expressive neural networks and the policy as a Gaussian with mean and covariance given by neural networks. We will next derive update rules for these parameter vectors.

\paragraph{The critic network} The Q-function parameters can be trained to minimize the Bellman residual.
\begin{equation}
\mathscr{L}\left(\phi \right)=  \left( y_h - Q_{\psi}\left(o_h, a_g, a_h, \bar{A}_h \right) \right)^2
\end{equation}
with 
\begin{equation}
y=r_h +\gamma V_{\psi^{-}}^{\mathrm{MF}}\left(o_h^{\prime}\right)
\end{equation}
The mean field value function $V^{MF}(o^\prime_h )$ for household agent $i$ over the government's action $a_g$ is
\begin{equation}
    V^{MF}_{\psi^-}\left(o^\prime_h \right)= \mathbb{E}_{a^\prime_h,\bar{A}^\prime_h \sim \pi_h(o^\prime_h) } Q(o^\prime_h, a^\prime_g, a^\prime_h, \bar{A}_h^\prime)
\end{equation}
The mean action $\bar{a}_h^i$ of all $i$'s neighbors is first calculated by averaging the actions $a^j$ taken by $i$'s $N^i$ neighbors from policies $\pi_t(o_h^j)$ parametrized by their previous mean actions $\bar{a}^j_{-}$
\begin{equation}
\bar{a}^i=\frac{1}{N^i} \sum_j a^j, a^j \sim \pi_t\left(\cdot \mid o_h^j, \bar{a}_{-}^j\right)
\label{mean action}
\end{equation}
The gradient of the households' critic network is given by
\begin{equation}
    \nabla_\psi \mathscr{L}\left(\phi \right) = \left( y_h - Q_{\psi}\left(o_h, a_g, a_h, \bar{A}_h \right) \right) \nabla_\psi Q_{\psi}\left(o_h, a_g, a_h, \bar{A}_h \right) 
\end{equation}

\paragraph{The actor network} 
With each $\bar{a}^i$ calculated as in Eq. \ref{mean action}, the policy network of household agents $\pi^{\sigma}_h$, i.e. the actor, of MF-AC is trained by the sampled policy gradient: 
\begin{equation}
\nabla_\sigma \mathscr{L}\left(\sigma \right) \approx  \log \pi_{\sigma}(o_h, \bar{A}_h)  Q_{\psi}\left(o_h, a_g, a_h, \bar{A}_h \right)
\end{equation}

\section{Additional Results}\label{additional results}

\subsection{Baselines Setting}\label{baseline setting}
\textbf{(1) Free Market Policy}
The government takes a hands-off approach to market activities, and households choose to work and saving actions stochastically.

\textbf{(2) Genetic Algorithm}
The government employs the genetic algorithm to optimize tax policies, while household agents adopt consumption and labor supply strategies following Heathcote's work~\cite{heathcote2014consumption}. See Appendix~\ref{GA} for more details.

\textbf{(3) Independent PPO}
Both the government and households utilize the PPO policy, treat other agents as part of the environment and learn policies based on local observations.

\textbf{(4) Multi-Agent Deep Deterministic Policy Gradient (MADDPG)}
To address large-scale multi-agent problems using MADDPG, we categorize the agents into 4 groups: the government, the top 10\% richest households, the top 10-50\% richest households, and the bottom 50\% richest households. Each group of agents shares an actor and a centralized critic and updates their networks using the MADDPG method.

\textbf{(5) Multi-Agent Proximal Policy Optimization (MAPPO)} 
Both government and household agents utilize a PPO actor and adopt a centralized value function. The government agent and household agents share a common actor network, with the observation and action spaces of heterogeneous agents aligned.

\textbf{(6) Bi-level Mean Field Actor-Critic (BMFAC)}
The government acts as the leader agent and updates its policy using the policy gradient method. While the household act as follower agents, sharing a common actor and critic, and updating networks via the mean-field actor-critic approach~\cite{yangMeanFieldMultiagent2018a}. The government and households utilize a bi-level optimization approach, iteratively optimizing their policies to converge toward a Markov Perfect Equilibrium. More details are provided in the Appendix~\ref{BMFAC}.

\textbf{(7) Heterogeneous-Agent Reinforcement Learning (HARL)} HARL algorithms (HAPPO, HATRPO, HAA2C) employ heterogeneous actors and a joint advantage estimator, without the need for relying on restrictive parameter-sharing trick.

\subsection{Training Curves}
\begin{figure*}
    \centering
    \includegraphics[scale=0.4]{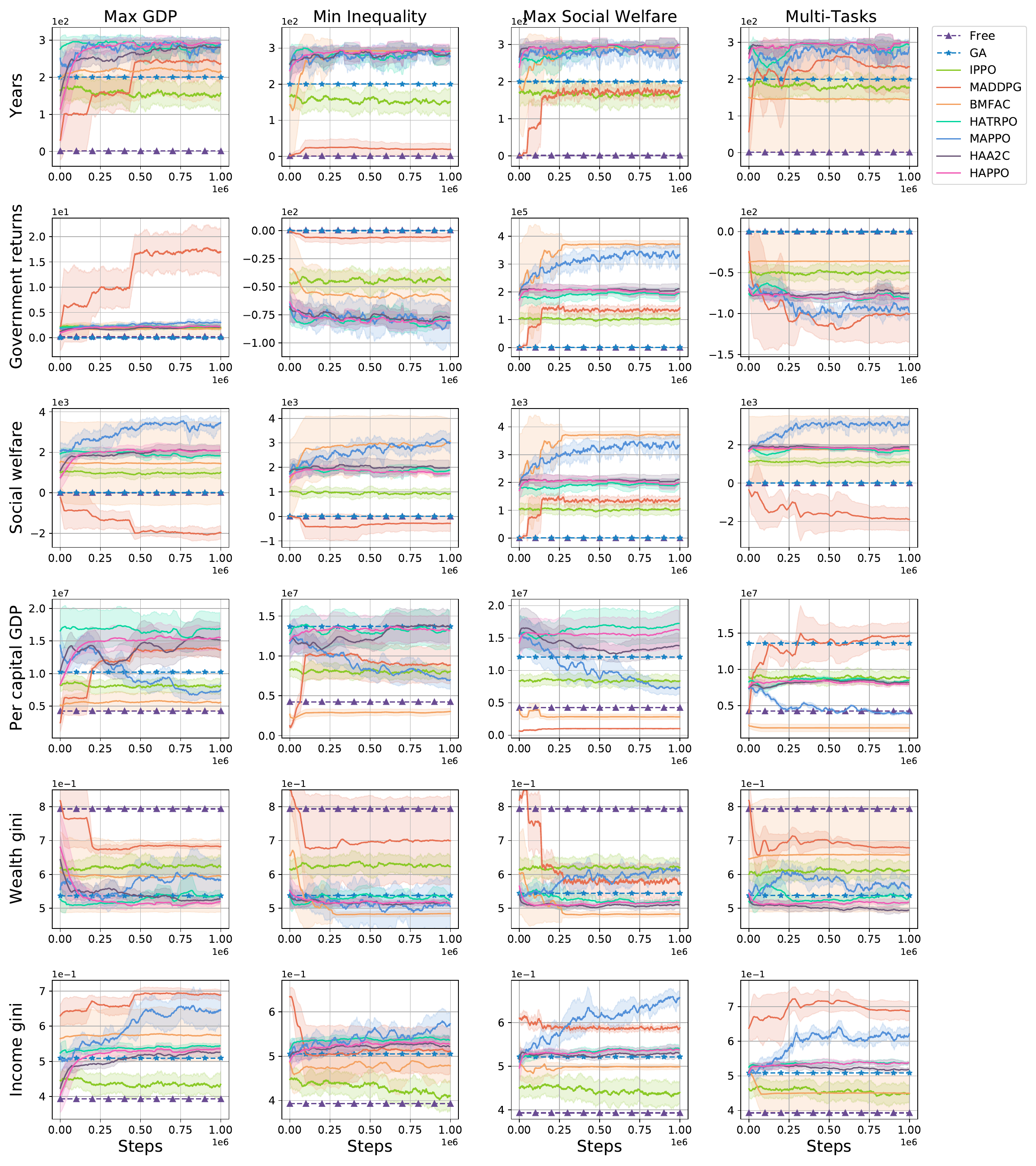}
    \caption{The training curves for 9 baselines on 6 economic indicators in 4 different tasks ($N=100$).}
    \label{fig:curves}
\end{figure*}

To showcase the learning performance of the MARL algorithms on the TaxAI simulator, we present training curves for 9 baselines across four different tasks (shown in Figure~\ref{fig:curves}). 
Among these baselines, GA and free-market policy are not learning algorithms, thus their convergence results are represented by straight lines. The MARL algorithms are represented by solid lines indicating the mean values, while the shaded areas reflect the variances of these economic indicators. In the subplots, each row represents an economic indicator, while each column corresponds to the experimental task.
From Figure~\ref{fig:curves}, it is evident that the MARL algorithm outperforms the traditional methods on six economic indicators across each task. Different algorithms excel in different tasks. For instance, HATRPO achieves the highest per capita GDP in the maximizing GDP task, while BMFAC achieves the lowest wealth Gini coefficient and the highest social welfare in the tasks of minimizing social inequality and maximizing welfare. In the multi-task scenario, different algorithms prioritize different optimization objectives. These results highlight the effectiveness and superiority of the MARL algorithms in finding optimal tax policies for the government and optimal working and saving strategies for households.

\subsection{Economic Evolution}\label{optimal solution}
To explore the optimal solution to the dynamic game between the government and households (N=100), we implement five baselines over three tasks: maximizing GDP growth rate, minimizing social inequality, and maximizing social welfare. Additionally, we test the converged policies and showcase the macroeconomic indicators, such as per capita GDP, wealth distribution, and social welfare, as the number of steps increases. This analysis aims to illustrate the evolutionary process of the economy under different algorithms.
\begin{figure*}
    \centering
    \includegraphics[scale=0.3]{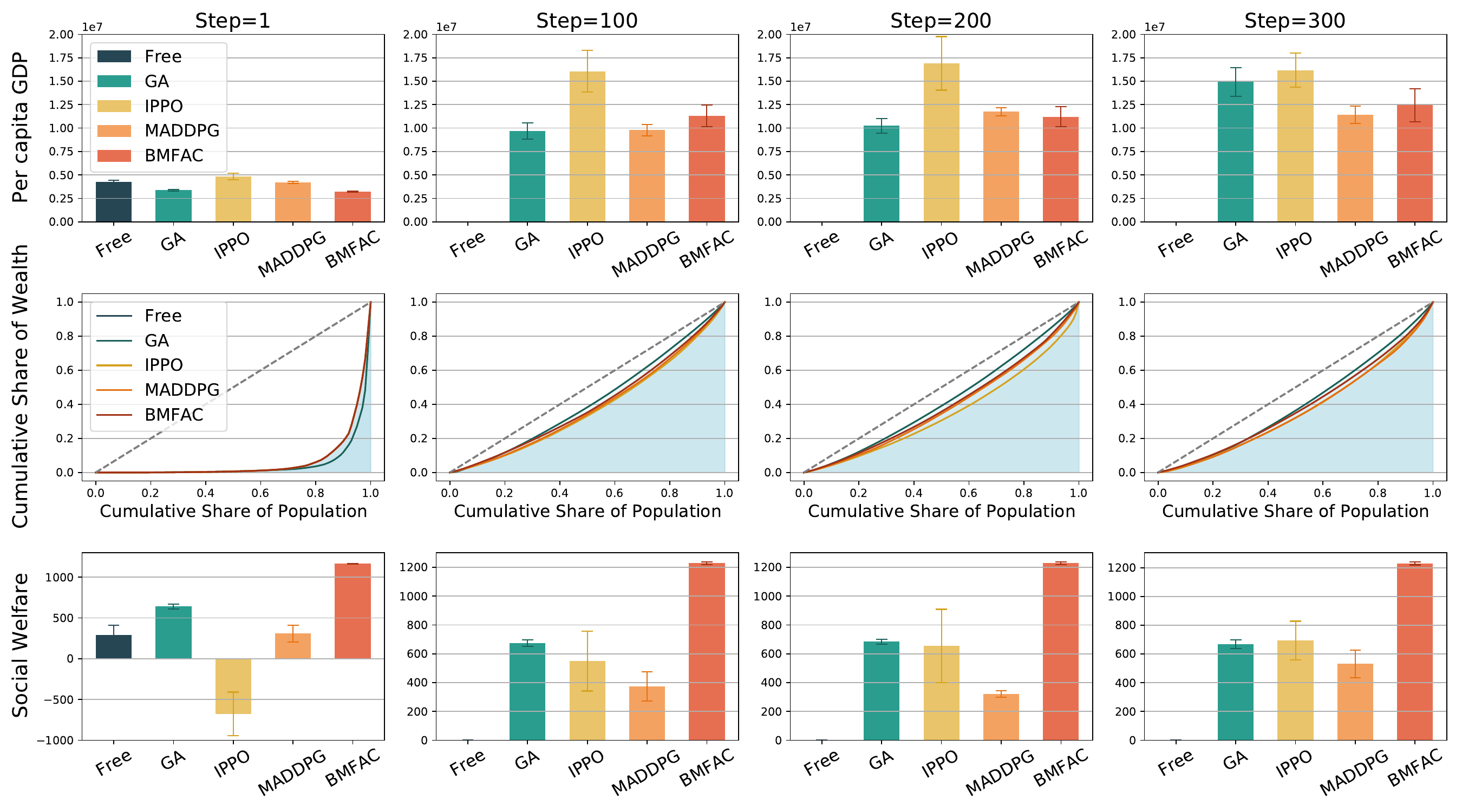}
    \caption{The economic evolution process of different algorithms in the TaxAI environment. Each row represents a macroeconomic indicator, while each column represents a time step in the evolution.}
    \label{fig:result}
\end{figure*}
In Figure~\ref{fig:result}, each row represents a macroeconomic indicator, with the first row corresponding to per capita GDP and the third-row representing social welfare. The horizontal axis represents different baselines and the height of the bars shows the GDP and social welfare at the current step. The second row displays the Lorenz curves for wealth distribution, where the vertical axis represents the cumulative proportion of wealth and the horizontal axis represents the cumulative proportion of a population ranked by their wealth. The five curves correspond to the Lorenz curves of different baselines, where a closer proximity to the line of perfect equality indicates a higher level of equality in wealth distribution.
Each column is divided based on steps, where we define each step as representing one year in the real world. The government and households make decisions at each time step. The maximum step is set to 300. When an economy satisfies a terminal condition, such as insufficient production to meet consumer demand or the Gini coefficient of wealth distribution exceeding a threshold, the economic system stops evolving.

The experimental results in Figure~\ref{fig:result} provide evidence for the ability of TaxAI to simulate the economic evolution and the feasibility and superiority of MARL algorithms in solving the optimal taxation problem.
Firstly, the free market policy struggles to progress beyond 100 steps, whereas the other four tax optimization policies can evolve up to 300 steps, thereby highlighting the efficacy of the proposed policy interventions for ensuring the sustainability of the economy. Furthermore, with the increasing steps, the GDP and social welfare associated with the GA, IPPO, MADDPG, and BMFAC methods continuously increase, while wealth inequality decreases. This alignment with the given optimization objectives reveals the feasibility of these methods in solving the optimal tax problem. Lastly, in the tasks evaluating GDP and social welfare, IPPO achieves the highest GDP, while BMFAC significantly outperforms other algorithms in terms of social welfare. In terms of the Lorenz curve, 5 baselines except the free market policy converge from an unequal state to a relatively higher level of equality. These results show that MARL methods possess the capability to simulate the evolution of the economy and attain optimal tax policies that surpass those achieved by traditional approaches such as the free market and GA.

The hyperparameters used in baselines algorithms are shown in Table~\ref{table: alg_hyperparameter}, \ref{table: IPPO}, \ref{table: MADDPG}.

\begin{table*}[ht]
\caption{Public Hyperparameters of Baselines Algorithms}
\label{table: alg_hyperparameter}
\vskip -0.5in
\begin{center}
\begin{small}
\begin{tabular}{l r}
\toprule
Hyperparameter & {Value} \\  
\midrule
{Discount factor $\gamma$}  & 0.975 \\
{Critic learning rate} & 3e-4 \\
{Actor learning rate}  & 3e-4 \\
{Replay buffer size}   & 1e6   \\
{Init exploration steps}  & 1000 \\
{Num of epochs }  & 1500  \\
{Epoch length}  &  500  \\
{Batch size}  &  128  \\
{Tau $\tau$}  &  5e-3   \\
{Evaluation epochs}  & 10  \\
{Hidden size} & 128 \\
\bottomrule
\end{tabular}
\end{small}
\end{center}
\end{table*}

\begin{table*}[h]
\caption{Hyperparameters of IPPO Algorithm}
\label{table: IPPO}
\begin{tabular}{l r}
\toprule
Hyperparameter & {Value} \\  
\midrule
  {Tau $\tau$}&0.95\\
  {Gamma $\gamma$}&0.99\\
  {Eps $\epsilon$}&1e-5\\
  {Clip}&0.1 \\
  {Vloss coef}&0.5 \\
  {Ent coef} &0.01 \\
  {Max grad norm}&0.5 \\
  {Update frequency} & 1  \\
\bottomrule
\end{tabular}
\end{table*}

\begin{table*}[h]
\caption{Hyperparameters of MADDPG Algorithm}
\label{table: MADDPG}
\begin{center}
\begin{small}
\begin{tabular}{l r}
\toprule
Hyperparameter & {Value} \\  
\midrule
{Update frequency} & 10  \\
{Noise rate }&0.1 \\
  {Epsilon} &0.1 \\
  {Save interval}&100\\
\bottomrule
\end{tabular}
\end{small}
\end{center}
\end{table*}

\begin{table*}[h]
\caption{Hyperparameters of Genetic Algorithm}
\label{table: ga hyperparameter}
\begin{center}
\begin{small}
\begin{tabular}{l r}
\toprule
Hyperparameter & {Value} \\  
\midrule
{DNA size}  & 12 \\
{Pop size} & 100 \\
{Crossover rate}  & 0.8 \\
{Mutation rate}   & 0.1   \\
{Max generations} & 200 \\
\bottomrule
\end{tabular}
\end{small}
\end{center}
\end{table*}

\begin{table*}[h]
\caption{Hyperparameters of MAPPO, HAPPO, HATRPO, HAA2C Algorithm}
\label{table: HARL}
\begin{center}
\begin{small}
\begin{tabular}{cc|cc|cc}
\toprule
 Hyperparameter & value & Hyperparameter & value & Hyperparameters & Value \\
\midrule
Eval episode & 20 & Optimizer & Adam & Num mini-batch & 1 \\
Gamma & 0.99 & Optim eps & $1 \mathrm{e}-5$ & Batch size & 4000 \\
Gain & 0.01 & Hidden layer & 1 & Training threads & 20 \\
Std y coef & 0.5 & Actor network & MLP & Rollout threads & 4 \\
Std x coef & 1 & Max grad norm & 10 & Episode length & 300 \\
Activation & ReLU & Hidden layer dim & 64 & Critic lr & $5 \mathrm{e}-3$ \\
Actor lr & $5 \mathrm{e}-5$ & Gov actor lr & $5 \mathrm{e}-4$ & Gov critic lr & $5 \mathrm{e}-4$ \\
\bottomrule
\end{tabular}
\end{small}
\end{center}
\end{table*}


\end{document}